\documentclass[prc,twocolumn,preprintnumbers,showpacs]{revtex4}
\usepackage{gensymb}
\usepackage{amsmath,amssymb,epsfig,bm,mathrsfs,color}
\usepackage{slashed}
\usepackage{color}
\newcommand{\be}{\begin{equation}}
\newcommand{\ee}{\end{equation}}
\newcommand{\bea}{\begin{eqnarray}}
\newcommand{\eea}{\end{eqnarray}}

\newcommand{\vecp}{{\bm p}}

\newcommand{\veck}{{\bm k}}
\newcommand{\vecQ}{{\bm Q}}
\newcommand{\vecR}{{\bm R}}
\newcommand{\vecr}{\bm r}

\newcommand{\SD}{^3S_1$-$^3D_1}

\definecolor{red}{rgb}{0.8,0,0}
\definecolor{orange}{rgb}{0.8,0.2,0.0}
\definecolor{blue}{rgb}{0.3,0.0,0.8}
\definecolor{green}{rgb}{0,0.5,0.0}
\definecolor{darkred}{rgb}{0.7,.1,.2}
\definecolor{bgred}{rgb}{1.,.95,.95}
\definecolor{bgblue}{rgb}{.95,.95,1.}
\definecolor{bluegreen}{rgb}{0.,.5,.3}
\definecolor{darkred}{rgb}{0.7,.1,.2}
\definecolor{darkgreen}{rgb}{0.1,.6,.0}
\definecolor{lightyellow}{rgb}{1.,1.,.8}
\definecolor{darkcyan}{rgb}{0.,.7,.9}
\definecolor{lightblue}{rgb}{0.6,0.8,1}
\definecolor{lightgreen}{rgb}{0.7,1.,.9}
\definecolor{money}{rgb}{0.4,0.8,0.}
\definecolor{purple}{rgb}{0.9,0.0,0.8}
\definecolor{orange}{rgb}{0.9,0.5,0.0}
\definecolor{newgr}{rgb}{0.2,0.8,0.2}
\definecolor{newbl}{rgb}{0.3,0.6,0.8}
\definecolor{newor}{rgb}{1.0,0.6,0.}

\begin{document}

\title{ BCS-BEC crossovers and unconventional phases in dilute nuclear  matter}

\preprint{INT-PUB-14-042}

\author{Martin Stein}
\email{mstein@th.physik.uni-frankfurt.de}
\affiliation{Institute for Theoretical Physics,
  J.~W.~Goethe-University, \\D-60438  Frankfurt am Main, Germany}

\author{ Armen Sedrakian}
\email{sedrakian@th.physik.uni-frankfurt.de}
\affiliation{Institute for Theoretical Physics,
  J.~W.~Goethe-University, \\D-60438  Frankfurt am Main, Germany}

\author{Xu-Guang Huang}
\email{huangxuguang@fudan.edu.cn}
\affiliation{Physics Department \& Center for Particle Physics and
  Field Theory, \\Fudan University, Shanghai 200433, China}

\author{John W. Clark}
\email{jwc@wuphys.wustl.edu}
\affiliation{Department of Physics and McDonnell Center for the Space 
Sciences, Washington University, St.~Louis, Missouri 63130, USA}
\affiliation{Centro de Ci\^encias Matem\'aticas, University of
Madeira, 9000-390 Funchal, Portugal
}

\date{\today}

\begin{abstract}
  We study the phase diagram of isospin-asymmetrical nuclear matter in
  the density-temperature plane, allowing for four competing phases of
  nuclear matter: (i) the unpaired phase, (ii) the translationally and
  rotationally symmetric, but isospin-asymmetrical BCS condensate,
  (iii) the current-carrying Larkin-Ovchinnikov-Fulde-Ferrell phase,
  and (iv) the heterogeneous phase-separated phase. The phase diagram
  of nuclear matter composed of these phases features two tri-critical
  points in general, as well as crossovers from the asymmetrical BCS
  phase to a BEC of deuterons plus a neutron gas, both for the
  homogeneous superfluid phase (at high temperatures) and for the
  heterogeneous phase (at low temperatures).  The BCS-BEC type
  crossover in the condensate occurs as the density is reduced.  We
  analyze in detail some intrinsic properties of these phases,
  including the Cooper-pair wave function, the coherence length, the
  occupation numbers of majority and minority nucleonic components,
  and the dispersion relations of quasiparticle excitations about the
  ground state. We show by explicit examples that the physics of the
  individual phases and the transition from weak to strong coupling
  can be well understood by tracing the behavior of these quantities.
\end{abstract}

\maketitle

\section{Introduction}

The two-nucleon vacuum interactions at low energies are well
constrained by the phase-shift data derived from the analysis of
elastic nucleon-nucleon collisions. Therefore, the main theoretical
challenge of understanding nuclear matter at sub-saturation densities
stems from the complexity of the many-body physics.  The attractive
part of the nuclear interaction is responsible for the formation of
nuclear clusters, as well as condensates of Bardeen-Cooper-Schrieffer
(BCS) type at low temperatures.  The temperature, density, and isospin
asymmetry of such matter are relevant for the description of
supernovae and neutron stars.  These two astrophysical venues differ
somewhat in the respective ranges of these variables.
For example, in supernovae the isospin asymmetries are much smaller
than in cold $\beta$-catalyzed neutron-star matter.  Consequently in
neutron-star matter $^1S_0$ pairing in the isospin-triplet,
spin-singlet state of neutrons is favored, whereas nearly
isospin-symmetrical matter supports $\SD$ pairing in the spin-triplet,
isospin-singlet state.

Fermionic BCS superfluids, which form loosely bound Cooper pairs at
weak coupling, undergo a transition to the Bose-Einstein condensate
(BEC) state of tightly bound bosonic dimers, once the pairing strength
increases sufficiently~\cite{1985JLTP...59..195N,1969PhRv..186..456E}.
This behavior has been confirmed in experiments on cold atomic gases,
where the interactions can be manipulated via the Feshbach mechanism.
In isospin-symmetric nuclear matter, the transition from the BCS to
the BEC state of the $\SD$ condensate may occur upon dilution of the
system, in which case the asymptotic state is a Bose-Einstein
condensate of
deuterons~\cite{1993NuPhA.551...45A,1995PhRvC..52..975B,1995ZPhyA.351..295S,2001PhRvC..63c8201L,
  2006PhRvC..73c5803S,2009PhRvC..79c4304M,2010PhRvC..81c4007H,2010PhRvC..82b4911J,2013JPhCS.413a2024S,2014JPhCS.496a2008S,2013PhRvC..88c4314P,2013PhRvC..88b5806S,2013arXiv1308.0364S,2013NuPhA.909....8S}.
Isospin asymmetry, induced by weak interactions in stellar
environments and expected in exotic nuclei, disrupts isoscalar
neutron-proton ($np$) pairing, because the mismatch in the Fermi
surfaces of protons and neutrons suppresses the pairing
correlations~\cite{2000PhRvL..84..602S}.  The standard
Nozi\`eres-Scmitt-Rink theory~\cite{1985JLTP...59..195N} of the
BCS-BEC crossover must also be modified, such that the low-density
asymptotic state becomes a gaseous mixture of neutrons and
deuterons~\cite{2001PhRvC..64f4314L}.  The $\SD$ condensates can be
important in a number of physical settings.  (i)~Low-energy heavy-ion
collisions produce large amounts of deuterons in final states as
putative fingerprints of $\SD$
condensation~\cite{1995PhRvC..52..975B}. (ii)~Large nuclei may feature
spin-aligned $np$ pairs, as evidenced by recent experimental
findings~\cite{2011Natur.469...68C} on excited states in $^{92}$Pd;
moreover, exotic nuclei with extended halos provide a locus for
$n$-$p$ Cooper pairing. (iii)~Directly relevant to the parameter
ranges covered in the present study are the observations that
supernova and hot proto-neutron-star matter at sub-saturation
densities have low temperature and low isospin asymmetry, and that the
deuteron fluid is a substantial
constituent~\cite{2010PhRvC..81a5803T,2009PhRvC..80a5805H}.

Two relevant energy scales for the problem domain under study are
provided by the magnitude of the shifts $\pm \delta\mu = \pm (\mu_n -
\mu_p)/2$ of the chemical potentials $\mu_n$ and $\mu_p$ of neutrons
and protons from their common value $\bar\mu$, and the pairing gap
$\Delta_0$ in the $\SD$ channel at $\delta\mu=0$.  With increasing
isospin asymmetry, i.e., as $\delta\mu$ increases from zero to values
of order $\Delta_0$, a sequence of unconventional phases may emerge.
One of these is a neutron-proton condensate whose Cooper pairs have
nonzero center-of-mass (c. m.) momentum
\cite{2001PhRvC..63b5801S,2003PhRvC..67a5802M,2009PhRvC..79c4304M};
this phase is the analog of the Larkin-Ovchinnikov-Fulde-Ferrell
(LOFF) phase in electronic
superconductors~\cite{LO,1964PhRv..135..550F}. Another possibility is
phase separation (PS) into superconducting and normal components,
proposed in the context of cold atomic
gases~\cite{2003PhRvL..91x7002B}. An alternative to the LOFF phase is
the deformed Fermi surface (DFS) phase, which, unlike the LOFF phase,
is translationally invariant but breaks the rotational
symmetry~\cite{2002PhRvL..88y2503M,2003PhRvC..67a5802M}.  Because these
two phases share many common properties, we shall concentrate only on
the LOFF phase. At large isospin asymmetry, where $\SD$ pairing is
strongly suppressed, a BCS-BEC crossover may also occur in the
isotriplet $^1S_0$ pairing channel, notably in neutron-rich systems
and halo
nuclei~\cite{PhysRevC.73.044309,2007PhRvC..76f4316M,2008PhRvC..78a4306I,PhysRevC.79.054305,2009PhRvC..79e4002A,2010PhLB..683..134S,2011PhRvC..84f7301S,2012PhRvC..86a4305S}. As
inferred from the experimental phase shifts, the pairing force in the
$\SD$ channel is stronger than in the $^1S_0$
channel. Isotriplet-spin-triplet 
pairing is prohibited by the Pauli principle;
accordingly, isotriplet pairing occurs only in the spin-singlet
channel. Because isosinglet-spin-triplet pairing is favored over
isotriplet spin-singlet pairing for not very high asymmetries, we
neglect isotriplet pairing. For large asymmetries, isosinglet
pairing is strongly suppressed and pairing takes place mostly in the
isotriplet spin-singlet channel.  Simple $^1S_0$ pairing only occurs
for chemical potentials in the continuum of two-particle scattering
states. However, pairing in the $\SD$ channel can arise for values of the
chemical potentials below the continuum edge, which is the case that
corresponds to bound states (deuterons).

In the first paper  (I) of this series~\cite{Stein:2012wd}, the
concepts of unconventional $\SD$ pairing and the BCS-BEC crossover
were unified in a model of isospin-asymmetrical nuclear matter by
including some of the phases mentioned above.  A phase diagram for
superfluid nuclear matter was constructed over wide ranges of density,
temperature, and isospin asymmetry. The coupled equations for the gap
and the densities of the constituents (neutrons and protons) were
solved allowing for the ordinary BCS state, its low-density asymptotic
counterpart BEC state, and two phases that owe their existence to the
isospin asymmetry: the phase with a current-carrying condensate (LOFF
phase) and the phase in which the normal fluid and superfluid occupy
separate spatial domains.  The latter phase is referred to as the
phase-separated BCS (PS-BCS) phase and, in the strong-coupling regime,
the phase-separated BEC (PS-BEC) phase. In this phase the asymmetry is
accumulated in the normal domains, whereas the superfluid domain is
perfectly isospin symmetric.

While the basic parameters of the superfluid phases, such as the
pairing gap and energy density have been studied extensively across
the BCS-BEC crossover, as well as in unconventional phases such as the
LOFF phase, some {\it intrinsic features} characterizing the
condensate are less well known.  These include the Cooper-pair wave
function, the occupation probabilities of particles, the coherence
length, and related quantities.  However, an understanding of the
evolution of these properties during the transitions from BCS to
unconventional (LOFF) phases as well as from weak to strong coupling
provide important insights into the mechanisms underlying the
emergence of new phases as well as into their nature.  The present
paper reports results from a study of these aspects of the pairing
problem for the example of the $\SD$ condensate carried out within the
framework developed in our previous work~\cite{Stein:2012wd}.  The
LOFF phase is chosen as a representative of the unconventional phases.
If PS takes place, one of the phases involved is the
isospin-symmetrical BCS phase, whereas the other is the normal
isospin-asymmetrical phase. Therefore, the intrinsic features of the
superfluid component of this phase, as specified above, are identical
to those of the BCS phase. Hence we do not discuss the intrinsic
properties of the PS-BCS phase.

To induce a BCS-BEC crossover in the condensate properties, we use as
a control parameter the adjustable density of the system.  The
relevant energies for scattering of two nucleons in the medium are set
essentially by their Fermi energies and in turn by the density of the
medium; hence the nuclear interaction strengths change with density as
well.

Accordingly, the BCS-BEC crossover is enforced by two effects: a
progressive dilution of the system and a concomitant increase in the
interaction strength in the $\SD$-channel at the lower energies
involved. In the present study, we additionally vary the isospin
asymmetry to generate a mismatch in the Fermi surfaces of paired
fermions, and we change the temperature to access the entire
density-temperature-asymmetry plane. It is worthwhile to note that in
ultracold atomic gases the BCS-BEC crossover is achieved in a
controlled manner by changing the effective interaction strengths via
the Feshbach mechanism, whereas the mismatch of Fermi surfaces is
achieved by trapping different amounts of atoms in a different
hyperfine states.

This paper is structured as follows. In Sec.~\ref{sec:theory} we give
a brief discussion of the theory of asymmetrical nuclear matter in the
language of imaginary-time finite-temperature Green's functions. In
Sec.~\ref{sec:phases} we discuss the phase diagram of asymmetrical
nuclear matter (Sec.~\ref{subsec:phase_diag}), the
temperature/asymmetry behavior of the gap in the weak-coupling regime
(Sec.~\ref{subsec:gap}), the kernel of the gap equation in BCS and
LOFF phases in various coupling regimes (Sec.~\ref{subsec:kernel}),
the Cooper-pair wave function across the BCS-BEC crossover
(Sec.~\ref{subsec:CooperWave}), and occupation numbers and
quasiparticle dispersion relations (Sec. \ref{subsec:numbers} and
\ref{subsec:DR}, respectively).  Our conclusions are summarized in
Sec.~\ref{sec:conclude}.

\section{Theory}
\label{sec:theory}

The Green's function of the superfluid, written in the Nambu-Gor'kov
basis, is given by 
\bea
\label{props} i\mathscr{G}_{12} =
i\left(\begin{array}{cc} G_{12}^{+} & F_{12}^{-}\\
    F_{12}^+ & G_{12}^{-}\end{array}\right) = \left(\begin{array}{cc}
    \langle T_\tau\psi_1\psi_2^+\rangle
    & \langle T_\tau\psi_1\psi_2\rangle \\
    \langle T_\tau\psi_1^+\psi_2^+\rangle & \langle T_\tau\psi_1^+\psi_2\rangle
\end{array}\right),
\eea 
where $G_{12}^{+}\equiv G^{+}_{\alpha\beta}(x_1,x_2)$, etc., $x =
(t,\vecr) $ denotes the continuous temporal-spatial variable, and
Greek indices label discrete spin and isospin variables.  Each
operator in Eq.~(\ref{props}) can be viewed as a bi-spinor, i.e.,
$\psi_{\alpha}=(\psi_{n\uparrow},\psi_{n\downarrow},\psi_{p\uparrow},\psi_{p\downarrow})^T,$
where the internal variables $\uparrow, \downarrow$ label a particle's
spin and the indices $n,p$ label its isospin.

The matrix {propagator} (\ref{props}) obeys the familiar Dyson
equation
\be \label{Dyson}
\left(\mathscr{G}_{0,13}^{-1}-\Xi_{13} \right) \mathscr{G}_{32} =
\delta_{12}, 
\ee 
where $\Xi_{12}$ is the matrix self-energy and the summation and
integration over repeated indices are implicit.
Equation~(\ref{Dyson}) can be transformed into momentum space, where
it becomes an algebraic equation.  For our purposes, translational
invariance cannot be assumed, so we proceed by defining relative
$\tilde r = x_1-x_2$ and c. m. $R = (x_1+x_2)/2$
coordinates and Fourier transforming with respect to the relative
four-coordinate and c. m. three-coordinate $\vecR$. The associated
relative momentum is denoted below by $k \equiv (ik_{\nu},\veck)$ and
the three-momentum of the c. m.is denoted by $\vecQ$. The zero component
of the vector $k$ takes on discrete values $k_{\nu} = (2\nu+1)\pi T $,
where $\nu\in \mathbb{Z}$ and $T$ is the temperature.

The relevant Fourier transformations can be obtained
by first performing a variable transformation to the 
c. m. and relative
coordinates
\begin{widetext}
\bea iG_{12}^{+}
=iG^{+}_{\tau\sigma,\tau'\sigma'}(\bm x_1,\bm x_2,\tilde t)
&=&\left\langle T\psi_{\tau\sigma}\left(\bm R+\frac{\tilde{\bm
        r}}2,0\right)\psi_{\tau'\sigma'}^+\left(\bm R-\frac{\tilde{\bm
        r}}2,\tilde t\right)\right\rangle ,
\eea
where to exploit the time translation invariance we 
have
defined the
relative time variable $\tilde t = t'-t$. The Fourier transformations
from the space-time to the 
momentum-frequency
domain are defined via 
\bea
G^{+}_{\tau\sigma,\tau'\sigma'}(\bm k,\bm Q,\tilde t)
=\frac1{(2\pi)^3}
\int d^3\bm R\,\, d^3\tilde{\bm r}\,\, e^{-i(\tilde{\bm r}\cdot\bm k+\bm R\cdot\bm Q)}
 G^{+}_{\tau\sigma,\tau'\sigma'}(\bm x_1,\bm x_2,\tilde t).
\eea
\end{widetext}
The Fourier transformation from the imaginary-time domain to the
frequency domain is given by 
\bea G^{+}_{\tau\sigma,\tau'\sigma'}(\bm k,\bm Q,t)
&=&\frac1\beta\sum_\nu e^{-ik_\nu
  t}G^{+}_{\tau\sigma,\tau'\sigma'}(ik_\nu,\bm k,\bm Q). \nonumber\\
\eea 
The Fourier transformations for the 
remaining
elements of the matrix
Green's function $i\mathscr{G}_{12}$ are constructed in an analogous
manner. With the definitions above, we obtain the Fourier image of
Eq.~(\ref{Dyson}) as
\be
\label{eq:Dyson2} 
\left[\mathscr{G}_0(k,\vecQ)^{-1}-\Xi(k,\vecQ) \right]\mathscr{G}
(k,\vecQ) = {\bf{1}}_{8\times 8}.  \ee 
Further reductions are possible owing to the fact that the normal
propagators for the particles and holes are diagonal in the
spin-isospin spaces, i.e., $(G^+,G^{-})\propto
\delta_{\alpha\alpha'}$, i.e., the off-diagonal elements of
$\mathscr{G}_0^{-1}$ are zero.  Writing out the non-vanishing
components in the Nambu-Gorkov space explicitly, we obtain
$[\mathscr{G}_0(ik_{\nu},\veck,\vecQ)^{-1}]_{11} =
-[\mathscr{G}_0(-ik_{\nu},\veck,-\vecQ)^{-1}]_{22}=
G_{0}^{-1}(ik_{\nu},\veck,\vecQ) $, where \be \label{inverseG}
{G}_0(k,\vecQ)^{-1} ={\rm diag}( ik_{\nu} - \epsilon_{n\uparrow}^+,
ik_{\nu} - \epsilon_{n\downarrow}^+, ik_{\nu} - \epsilon_{p\uparrow}^+
, ik_{\nu} - \epsilon_{p\downarrow}^+ ) \ee with 
\bea
\label{eq:norm_spec}
\epsilon^{\pm}_{n/p,\uparrow/\downarrow} &=& \frac{1}{2m^*}\left(\veck\pm
  \frac{\vecQ}{2}\right)^2-\mu_{n/p}, 
\eea
which it is useful to separate into sym\-met\-ri\-cal anti-symmetrical
parts with respect to the time-reversal operation by writing
\bea
\label{eq:spectra1}
\epsilon_{n\uparrow/\downarrow}^{\pm} &=&  E_S-\delta\mu\pm E_A,\\
\label{eq:spectra2}
\epsilon_{p\uparrow/\downarrow}^\pm &=& E_S+\delta\mu \pm E_A,
\eea 
where 
\bea
\label{eq:E_S}
E_S &=&\frac{Q^2/4+k^2}{2m^*}-\bar\mu, \\
\label{eq:E_A}
E_A &=& \frac{\veck\cdot \vecQ}{2m^*} ,
\eea 
are the symmetrical and anti-symmetrical parts of the quasiparticle
spectrum and $\bar\mu \equiv (\mu_n+\mu_p)/2$.  The effective mass
$m^*$ is defined in the usual fashion in terms of the normal
self-energy, bare mass $m$, and Fermi momentum $p_F$,
specifically
\be
m/m^* =
1-(m/p) \partial_p \Xi_{11} \vert_{p=p_F},
\ee
if we neglect the small mismatch between neutron and proton effective 
masses.

Keeping this mismatch implies the changes $E_{S/A}\to
E_{S/A}(1\pm\delta_m)$ and $\delta\mu\to \delta\mu+\mu\delta_m$, where
$\delta_m = (m^*_n-m^*_p)/(m^*_n+m^*_p)\ll 1$. In the analysis below,
$\delta_m$ lies in the range $0\le \vert \delta_m\vert \le 0.06$, the
upper bound being attained for the largest asymmetries and densities
relevant to this study.
The quasiparticle spectra in Eq.~(\ref{inverseG}) are
written in a general reference frame moving with the c. m. momentum
$\vecQ$ relative to a laboratory frame at rest.  The spectrum of
quasiparticles is seen to be two-fold degenerate; i.e., the SU(4)
Wigner symmetry of the unpaired state is broken down to spin
SU(2).  In fact this Wigner symmetry is always approximate, 
because the phase shifts in the isoscalar and isotriplet $S$ waves 
differ, such that isosinglet pairing is stronger than isotriplet 
pairing in bulk nuclear matter.

The nucleon-nucleon scattering data indicates that the dominant
attractive interaction in low-density nuclear matter is the $\SD$
partial wave, which leads to isoscalar (neutron-proton) spin-triplet
pairing.  Accordingly, the anomalous propagators have the property
$(F^+_{12},F^-_{12})\propto  (-i\tau_y)  
\otimes
\sigma_x 
$, where
$\sigma_i$ and $\tau_i$ are Pauli matrices in spin and isospin spaces.
This implies that in the quasiparticle approximation, the self-energy
$\Xi$ has only off-diagonal elements in the Nambu-Gorkov space.
Specifically, $\Xi_{12} = \Xi_{21}^{+} = i\Delta_{\alpha\beta}$, with
$\Delta_{14}= \Delta_{23} =-\Delta_{32} = - \Delta_{41} \equiv
\Delta$, where $\Delta$ is the (scalar) pairing gap in the $\SD$
channel.  Thus the first multiplier on the left-hand-side of
Eq.~(\ref{eq:Dyson2}) is given by
\begin{widetext}
\begin{eqnarray}
\mathscr{G}_0^{-1}-\Xi &=&\begin{pmatrix}
   ik_{\nu} - \epsilon_{n\uparrow}^+&0&0&0 &0&0&0&i\Delta\\ 0&ik_{\nu}- \epsilon_{n\downarrow}^+&0&0 &0&0&i\Delta&0\\ 0&0& ik_{\nu} - \epsilon_{p\uparrow}^+ &0 &0&-i\Delta&0&0\\ 0&0&0&ik_{\nu} - \epsilon_{p\downarrow}^+ &-i\Delta&0&0&0\\
   0&0&0&i\Delta &ik_{\nu} + \epsilon_{n\uparrow}^-&0&0&0\\ 0&0&i\Delta&0  &0&ik_{\nu}+ \epsilon_{n\downarrow}^-&0&0\\  0&-i\Delta&0&0 &0&0& ik_{\nu} + \epsilon_{p\uparrow}^- &0\\ -i\Delta&0&0&0 & 0&0&0&ik_{\nu} + \epsilon_{p\downarrow}^-
 \end{pmatrix}.
\end{eqnarray}
It is sufficient to consider only a $4\times4$ block of the full
$8\times 8$ matrix Dyson equation, as the remaining blocks do not
contain new information. We consider then
\begin{eqnarray}
\begin{pmatrix}
    ik_{\nu} - \epsilon_{n}^+&0&0&i\Delta\\ 0& ik_{\nu} - \epsilon_{p}^+ &-i\Delta&0\\ 0&i\Delta &ik_{\nu} + \epsilon_{n}^-&0\\ -i\Delta&0&0&ik_{\nu} + \epsilon_{p}^-
  \end{pmatrix}\cdot
  \begin{pmatrix}
    \begin{matrix} G^+_n&0\\0&G^+_p \end{matrix} &
    \begin{matrix} 0&F^-_{np}\\F^-_{pn}&0 \end{matrix} \\
    \begin{matrix} 0&F^+_{np}\\F^+_{pn}&0 \end{matrix} &
    \begin{matrix} G^-_n&0\\0&G^-_p \end{matrix} &
  \end{pmatrix}=   {\rm diag}(1,1,1,1).
\end{eqnarray}
\end{widetext}
The solutions of this equation provide the normal and anomalous
Green's functions
\bea
G_{n/p}^{\pm} &=&
\frac{ik_{\nu}\pm\epsilon_{p/n}^{\mp}}{(ik_{\nu}-E^+_{\mp/\pm})(ik_{\nu}+E^-_{\pm/\mp})},\\
F_{np}^{\pm} &=&
\frac{-i\Delta}{(ik_{\nu}-E^+_{\pm})(ik_{\nu}+E^-_{\mp})},\\
F_{pn}^{\pm} &=&
\frac{i\Delta}{(ik_{\nu}-E^+_{\mp})(ik_{\nu}+E^-_{\pm})},
\eea
where the four branches of the quasiparticle 
spectrum are given by 
\be 
\label{eq:QP_spectra}
E_{r}^{a} = \sqrt{E_S^2+\Delta^2} + r\delta\mu +aE_A, 
\ee 
in which $a, r \in \{+,-\}$.  When $r=a$ and $ E_A>0$ the shifts owing to the isospin
asymmetry $\delta\mu$ and owing to the c. m. momentum $\vecQ$ add up;
consequently the branches $E_{-}^{-}$ and $E_{+}^{+}$ are located
farther away from the isospin-symmetrical spectrum than the branches
with $r\neq a$ for which these two factors compensate for each other.
In mean-field approximation, the anomalous self-energy (pairing-gap)
is determined by
\bea \label{eq:gap}
\Delta(\veck,\vecQ) &=&  \frac{1}{4\beta} \int\!\!\frac{d^3k'}{(2\pi)^3}\sum_{\nu}
V(\veck,\veck')               \nonumber\\
&& {\rm Im}  \Bigl[  
  F^+_{np} (k'_{\nu},\veck',\vecQ)
+F^-_{np} (k'_{\nu},\veck',\vecQ)  \nonumber\\
&-&F^+_{pn} (k'_{\nu},\veck',\vecQ) 
-F^+_{pn} (k'_{\nu},\veck',\vecQ) 
\Bigr],
\eea
where $ V(\veck,\veck')$ is the neutron-proton interaction potential.

We perform a partial-wave expansion in Eq.~(\ref{eq:gap}) and 
compute  the Matsubara sum, which yields 
\bea \label{eq:gap2}
\Delta_l(Q) &=& \frac{1}{4}\sum_{a,r,l'} \int\!\!\frac{d^3k'}{(2\pi)^3}
V_{l,l'}(k,k') 
\nonumber\\ &\times&
\frac{\Delta_{l'}(k',Q)}{2\sqrt{E_{S}^2(k')+\Delta^2(k',Q)}}[1-2f(E^r_a)],
\eea
where $V_{l,l'}(k,k')$ is the interaction in the $\SD$ partial wave,
$f(\omega)=1/[\exp{(\omega/T)}+1]$, and $\Delta^2 =\sum_l \Delta_l^2$.

The densities of neutrons and protons in any of the superfluid states
are obtained by observing that
\bea\label{eq:densities}
\rho_{n/p} (\vecQ)&=&\frac{2}{\beta}\int\!\!\frac{d^3k}{(2\pi)^3}\sum_{\nu} 
G^+_{n/p}(k_{\nu},\veck,\vecQ)\nonumber\\
&=&2 \int\!\!\frac{d^3k}{(2\pi)^3}
\Biggl[
\frac{1}{2}               \left(1+\frac{E_S}{\sqrt{E_S^2+\Delta^2}}\right) f(E^+_{\mp})\nonumber\\
&+&\frac{1}{2}             \left(1-\frac{E_S}{\sqrt{E_S^2+\Delta^2}}\right) f(-E^-_{\pm})
\Biggr].
\eea 
The magnitude $Q$ of the c. m. momentum in Eqs.~(\ref{eq:densities}) and
(\ref{eq:gap2}) is a parameter to be determined by minimizing
the free energy of the system.  For the homogeneous (but
possibly translationally noninvariant) cases it suffices to find
the minimum of the free energy of the superfluid ($S$) or unpaired ($N$) 
phase,
\be \label{eq:free}
F_S =  E_S-TS_S,\quad F_N =  E_N-TS_N,
\ee
where $E$ is the internal energy (statistical average of the system
Hamiltonian) and $S$ denotes the entropy.  The free energy of the
heterogeneous superfluid phase, which corresponds to separation of the
normal and superfluid phases, is constructed as a linear
combination,
\bea \label{eq:free_mixed}
\mathscr{F}(x,\alpha) = (1-x) F_S(\alpha = 0) 
+ x F_N(\alpha \neq 0), \,\, (Q = 0),\nonumber\\
\eea
where $x$ here denotes the filling fraction of the unpaired component
and \be \alpha = \frac{\rho_n-\rho_p}{\rho_n+\rho_p} \ee is the
density asymmetry.  In the superfluid phase (S) one has
$\rho_n^{(S)}=\rho_p^{(S)} = \rho^{(S)}/2$, while in the unpaired phase
(N) the neutron and proton partial densities are rescaled to new
values $\rho_{n/p}^{(N)}$. Thus, the net densities of neutrons/protons
per unit volume are given by $\rho_{n/p} = ({1}/{2})(1-x)\rho^{(S)} +
x\rho_{n/p}^{(N)}$.

The four possible states we consider are characterized as follows:
\bea\label{eq:phases}
\left\{
\begin{array}{llll}
Q = 0,  &\Delta \neq 0, & x = 0,& \textrm{BCS phase,}\\
Q \neq 0, & \Delta \neq 0, &x = 0,&\textrm{LOFF phase,} \\
Q = 0,  &\Delta \neq 0, & x \neq 0,&\textrm{PS phase,}\\
 Q = 0, & \Delta = 0, & x = 1, &\textrm{unpaired phase,} \\
\end{array}
\right. 
\eea
and we assign the ground state to the phase with lowest free energy at any
given temperature, density and isospin asymmetry.
\begin{figure}[tb]
\begin{center}
\includegraphics[width=8.5cm,height=7cm]{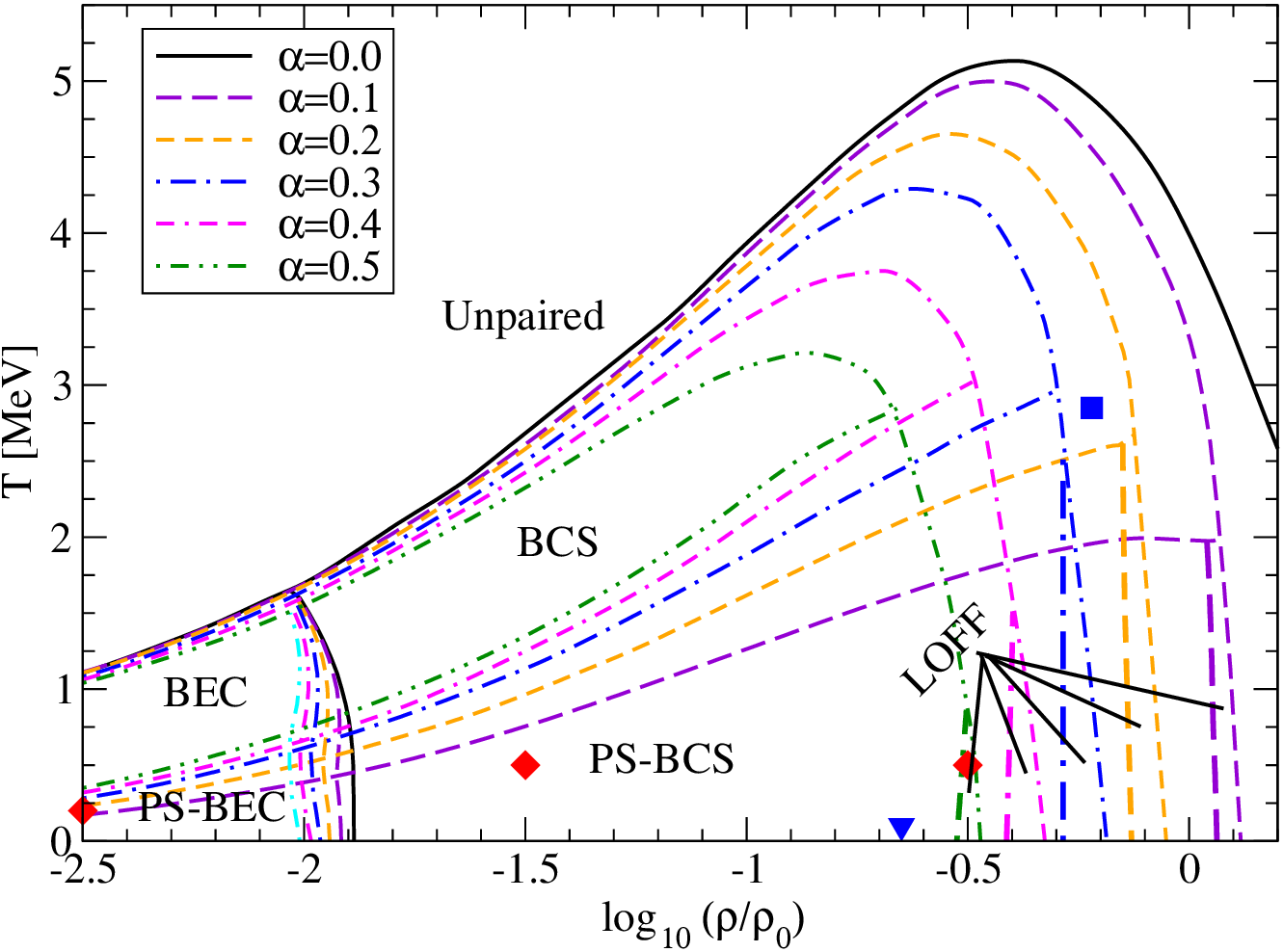}
\caption{(Color online) Phase diagram of dilute nuclear matter in the
  temperature-density plane for several isospin asymmetries $\alpha$
  (see also Ref.~\cite{Stein:2012wd}). Here $\rho_0 = 0.16$ fm$^{-3}$ is
  the nuclear saturation density for $\alpha =0$.  Included are four
  phases: unpaired phase, BCS (BEC) phase, LOFF phase, and PS-BCS
  (PS-BEC) phase. For each asymmetry there are two tri-critical
  points, one of which is always a Lifshitz
  point~\cite{1980JMMM...15..387H}.  For special values of asymmetry,
  these two points degenerate into a single tetra-critical point
 at $\log(\rho/\rho_0) = -0.22$ and $T = 2.85$ MeV when
  $\alpha_4 = 0.255$ (shown by a square).  The LOFF phase
  disappears at the point $\log(\rho/\rho_0) = -0.65$ and $T=0$ (shown
  by a triangle) for $\alpha = 0.62.$ The boundaries between BCS and
  BEC phases are identified by the change of sign of the average
  chemical potential $\bar\mu$. The diamonds (red online) mark the
  density and temperature values in the diagram that are used in the
  representative study (see Table I) of weak-coupling,
  intermediate-coupling, and strong-coupling regimes (from right to
  left).}
\label{fig:phasediagram}
\end{center}
\end{figure}
Inputs for the subsequent numerical calculations are the same as in
paper I.  Specifically, the pairing interaction is given by the bare
nucleon-nucleon interaction in the $\SD$ partial wave, based on the
(phase-shift equivalent) Paris potential \cite{1984PhRvC..30.1822H}.
The nuclear mean field is modeled by a Skyrme density functional, with
SkIII \cite{1987PhRvC..35.1539S} and SLy4 \cite{1998NuPhA.635..231C}
parametrizations yielding nearly identical results. A computation of
the effective mass from a realistic (e. g. Paris) potential would
require a larger numerical effort within a beyond-mean-field
microscopic many-body approach. However the effective masses computed
from microscopic approaches agree well with those derived from Skyrme
functionals and our results are not sensitive to small (of order
of a few percent) variations in the effective mass.

\section{BCS phase, LOFF phase, and crossover to BEC}
\label{sec:phases}

\subsection{Phase diagram}
\label{subsec:phase_diag}

The phase diagram of dilute nuclear matter is shown in
Fig.~\ref{fig:phasediagram} for several values of isospin asymmetry
$\alpha$.  Four different phases of matter are present [see
Eq.~\eqref{eq:phases}].  (a) The unpaired normal phase is always the
ground state at sufficiently high temperatures $T>T_{c0}$, where
$T_{c0}(\rho)$ is the critical temperature of the normal/superfluid
phase transition at $\alpha =0$. (b) The LOFF phase is the ground
state in a narrow temperature-density strip at low temperatures and
high densities. (c) The domain of PS appears at low
temperatures and low densities. Finally, (d) the isospin-asymmetrical
BCS phase is the ground state at intermediate temperatures and
intermediate to low densities.  It is convenient at this point to
define three regimes of coupling which are characterized solely by the
density of the system, because the boundaries between these regimes are
insensitive to the temperature. The strong-coupling regime (SCR)
corresponds to the low-density limit where well-defined deuterons are
formed, while the weak-coupling regime (WCR) corresponds to the
high-density limit where well-defined Cooper pairs are present.  In
between these limiting cases we identify the intermediate-coupling
regime (ICR). We  delineate the boundaries between these regimes in
the following discussion.

At the extreme of low density corresponding to the SCR, the BCS
superfluid phases have two counterparts: The BCS phase evolves into
the BEC phase of deuterons, whereas the PS-BCS phase evolves into the
PS-BEC phase, in which the superfluid fraction of matter is a BEC of
deuterons.  The superfluid/unpaired phase transitions and the phase
transitions between the superfluid phases are of second order (thin
solid lines in Fig.~\ref{fig:phasediagram}), with the exception of the
PS-BCS to LOFF transition, which is of first order (thick solid lines
in Fig.~\ref{fig:phasediagram}). The BCS-BEC transition and the PS-BCS
to PS-BEC transition are smooth crossovers.  At nonzero isospin
asymmetry, the phase diagram features two tri-critical points, i.e.,
points where the simpler pairwise phase coexistence terminates and
three different phases coexist.

Consistent with the earlier studies of the BCS-BEC crossover, one
observes in the phase diagram of Fig.~\ref{fig:phasediagram} a smooth
crossover to an asymptotic state corresponding to a mixture of a Bose
condensate of deuterons and a gas of excess neutrons. This however
occurs at moderate temperatures, where the unconventional phases do
not appear.  The new ingredient of the nuclear phase diagram is the
crossover seen at very low temperatures, where the heterogeneous
superfluid phase is replaced by a heterogeneous mixture of a phase
containing a deuteron condensate and a phase containing neutron-rich
unpaired nuclear matter.

\subsection{Temperature and asymmetry dependence of the gap:
 contrasting the BCS and LOFF phases}
\label{subsec:gap}

\begin{figure}[t]
\begin{center}
\includegraphics[width=8cm]{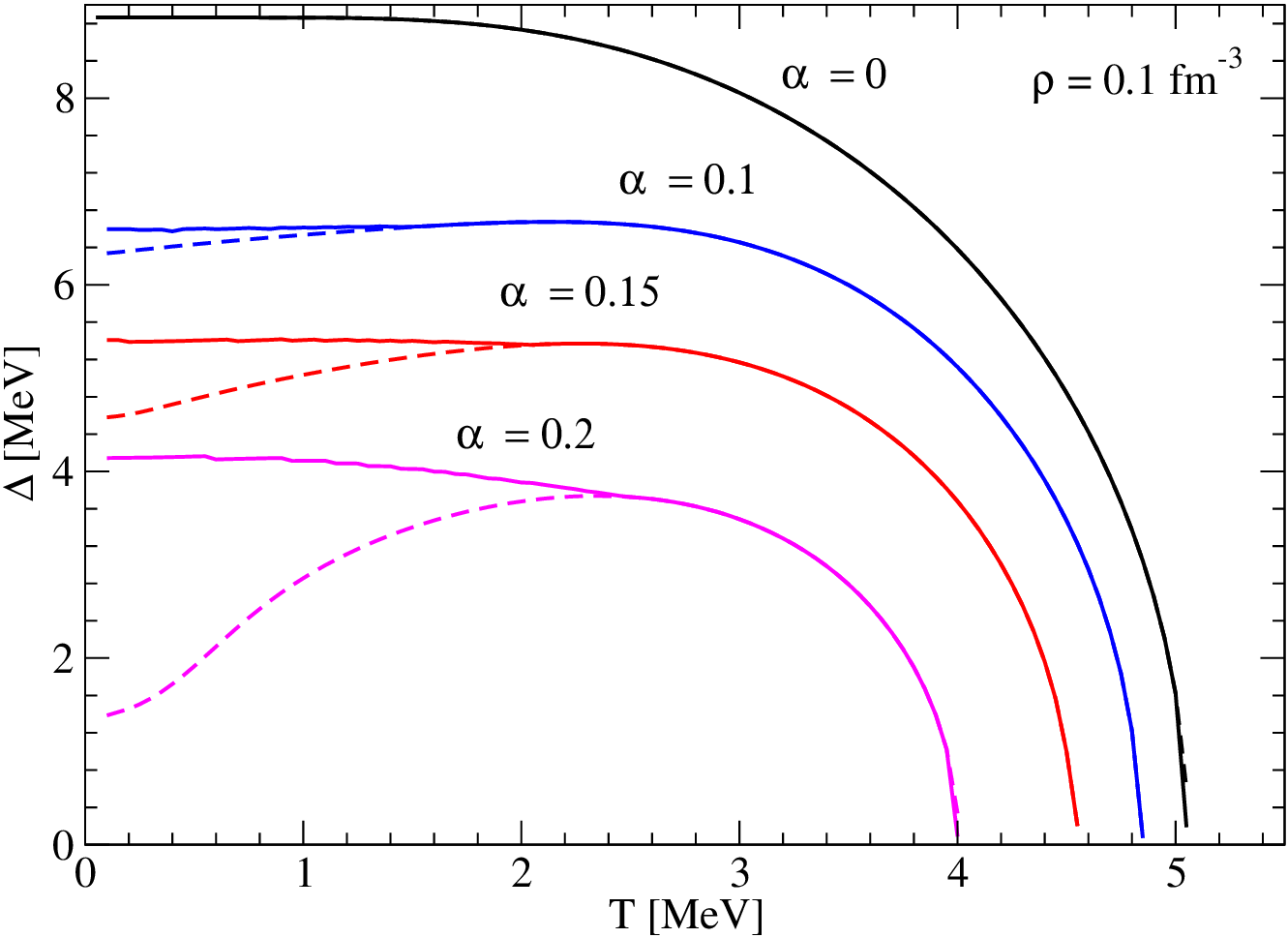}
\caption{(Color online) Gap as a function of temperature for asymmetry
  values $\alpha=0.0$ (black), $\alpha=0.1$ (blue),
  $\alpha=0.15$ (red) and $\alpha=0.2$ (magenta).  Results
  allowing for the LOFF phase are traced by solid lines, those
  restricted to the BCS phase are traced by dashed lines.}
\label{gap_loff}
\end{center}
\end{figure}

\begin{figure}[tb]
\begin{center}
\includegraphics[width=8cm]{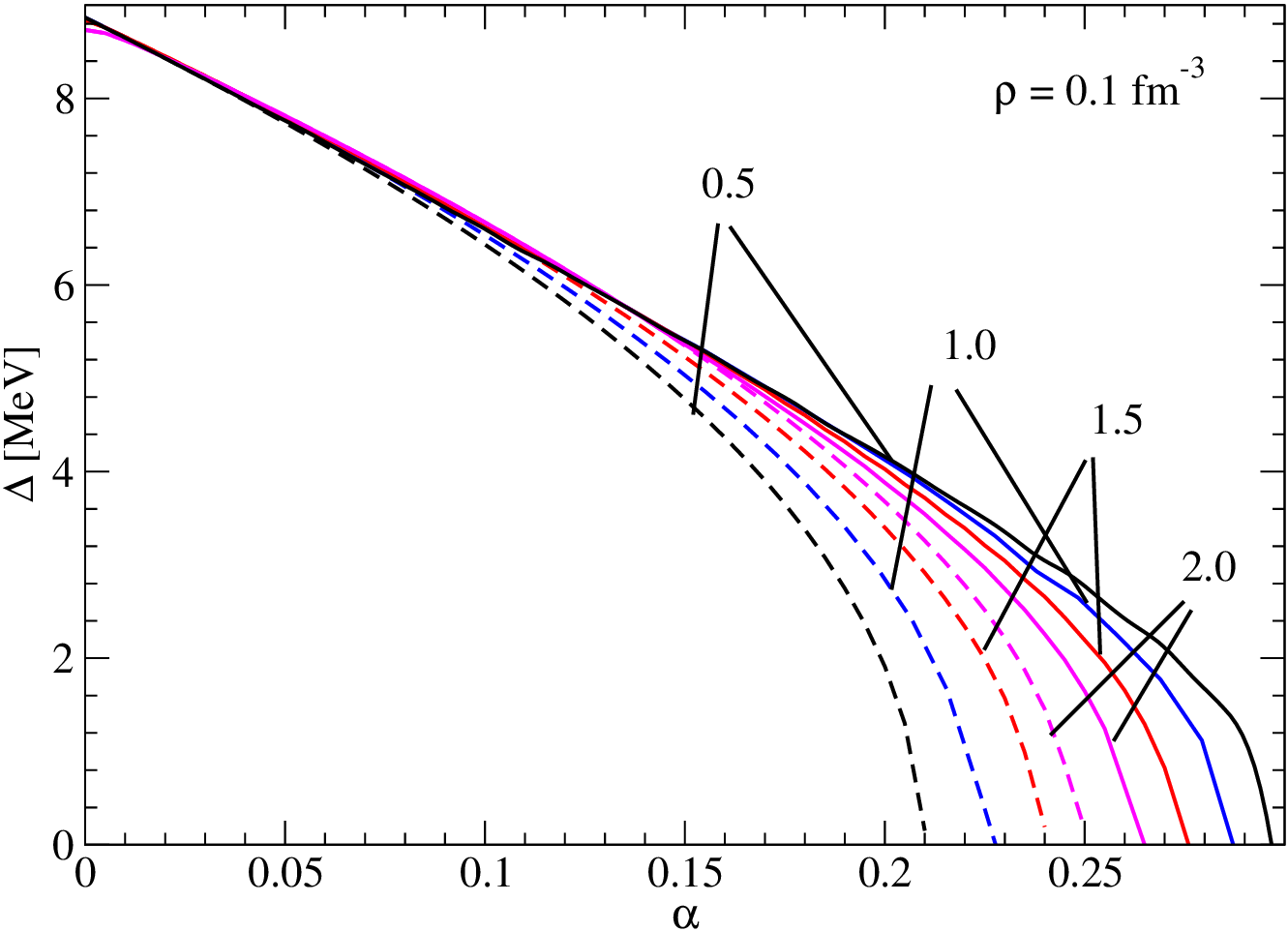}
\caption{(Color online) Gap as a function of asymmetry at constant
  density $\rho=0.1$ fm$^{-3}$ for $T=0.5$ (black), $T=1.0$ MeV
  (blue), $T=1.5$ MeV (red), $T=2.0$ MeV (magenta).  Results allowing
  for the LOFF phase are traced by solid lines; those restricted to
  the BCS phase are traced by dashed lines.}
\label{gap_alpha}
\end{center}
\end{figure}

Before turning to the main topic of this work, we would like to recall
and explore the behavior of the gap function as a function of
temperature and asymmetry at constant density. We concentrate only on
the WCR, as the behavior of the gap function in
SCR is self-similar to that of the WCR.  For
now, we also neglect the possibility that the PS phase is the ground
state.  Figure \ref{gap_loff} shows the weak-coupling gap as a
function of temperature for a range of asymmetries.  The plotted
results for each nonzero value of $\alpha$ reveal different regimes of
relatively low and relatively high temperature that reflect the
different behaviors of the gap when the possibility of a LOFF phase is
taken into account (solid curves) and when it is not (dashed curves).
Two branches existing at lower temperatures merge at some point to
form a single segment existing at higher temperatures.  This
high-temperature segment corresponds to the BCS state, and the
temperature dependence of the gap is standard, with $d\Delta(T)/dT <
0$ and asymptotic behavior 
$\Delta(\alpha, T) \sim [T_c(\alpha)(T_c(\alpha)-T)]^{1/2}$ as 
$T \to T_c(\alpha)$, where $T_c(\alpha)$ is the (upper) critical 
temperature.  In the
low-temperature region below the branch point,
there are two competing phases (BCS and LOFF), with very different 
temperature dependences of the gap function.  The quenching of the 
BCS gap (dashed lines) as the temperature is decreased is caused by 
the loss of coherence among the quasiparticles 
as the thermal smearing of the Fermi surfaces disappears.
Consequently, in the low-temperature range below the branch
point, the BCS branch shows the unorthodox behavior 
$d\Delta(T)/dT > 0$, and for large enough asymmetries there exists a
lower critical temperature $T_c^*$~\cite{2000PhRvL..84..602S}.  On the
contrary, one finds $d\Delta(T)/dT < 0$ for the LOFF branch, as is the
case in ordinary (symmetrical) BCS theory~\cite{2006PhRvB..74u4516H}.
It should be mentioned that the ``anomalous" behavior of the BCS gap
below the point of bifurcation leading to the LOFF state gives rise to
a number of anomalies in thermodynamic quantities, such as negative
superfluid density or excess entropy of the
superfluid~\cite{2006PhRvL..97n0404S}.  These anomalies are absent in
the LOFF state~\cite{2007IJMPE..16.2363J}.  Figure~\ref{gap_alpha}
shows the dependence of the gap function on asymmetry for several
pertinent temperatures.  In accord with Fig.~\ref{gap_loff}, there are
two curves (or segments) for each temperature: one in the low-$\alpha$
domain where only the BCS phase exists and the other in the
large-$\alpha$ domain where both BCS (dashed lines) and LOFF states
(solid lines) are possible. Clearly the LOFF solution, for which the
gap extends to larger $\alpha$
values, is favored in the latter domain.  

For small $\alpha$ the gap function is linear in $\alpha$.  At the
other extreme of large $\alpha$, the gap has the asymptotic behavior
$\Delta(\alpha)\sim \Delta_{00} \left(1-\alpha/\alpha_1\right)^{1/2}$,
where $\alpha_1\sim \Delta_{00}/\bar\mu$ and $\Delta_{00}$ is the
value of the gap at vanishing temperature and asymmetry.  The critical
asymmetry $\alpha_2$ at which the LOFF phase transforms into the
normal phase is a decreasing function of temperature, whereas that for
termination of the BCS phase (denoted $\alpha_1$ above) increases up
to the temperature where $\alpha_1=\alpha_2$.  For larger
temperatures, $\alpha_1$ decreases with temperature. Consequently, in
the dominant phase the critical asymmetry always decreases with
temperature.

\begin{figure}[tb]
\begin{center}
\includegraphics[width=8cm]{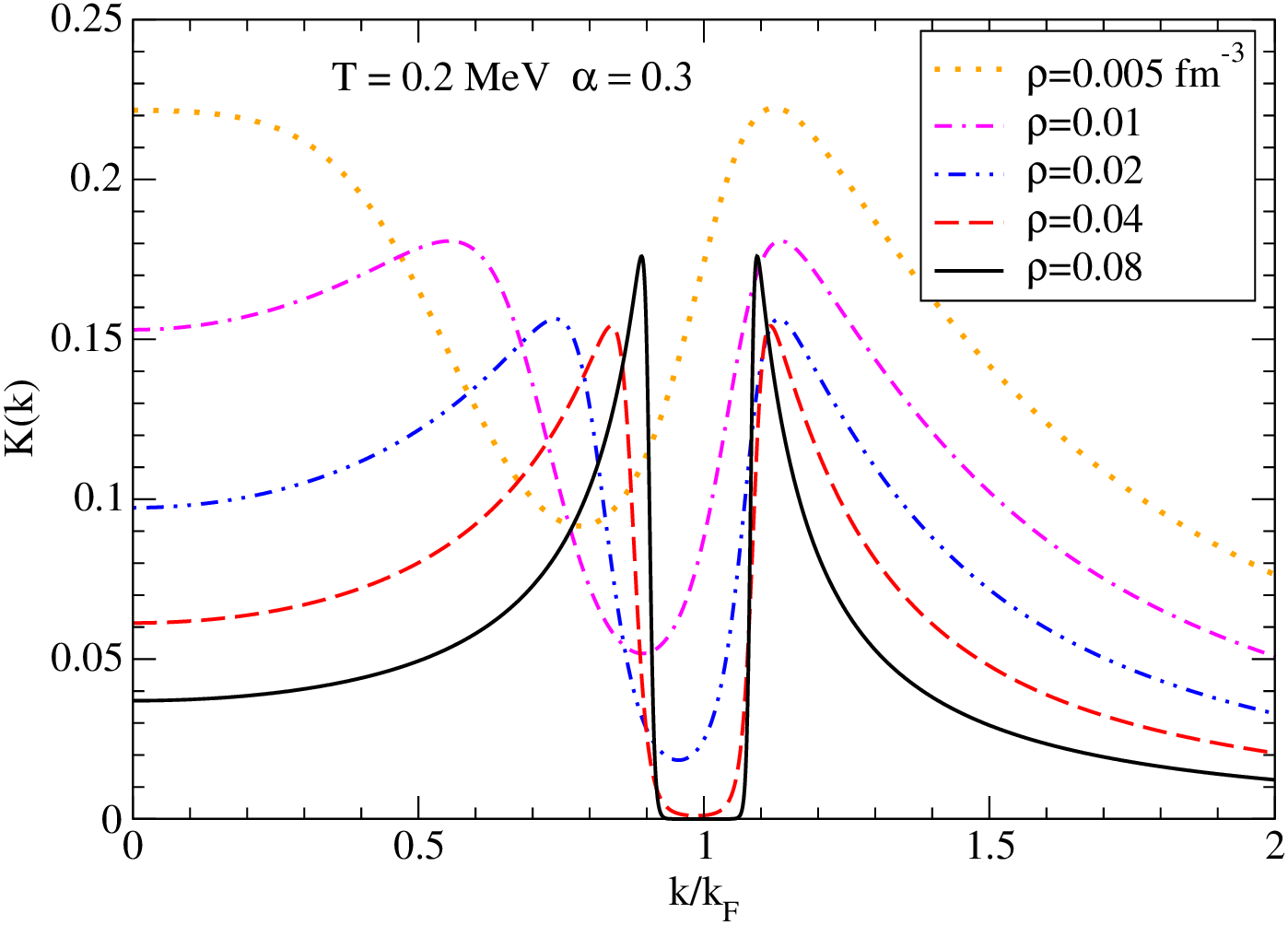}
\caption{(Color online) Dependence of the kernel $K(k)$ on momentum in
  units of Fermi momentum for fixed $T=0.2$ MeV, $\alpha = 0.3$, and
  various densities indicated in the plot.  }
\label{fig:kernel_density}
\end{center}
\end{figure}
\begin{figure}[tb]
\begin{center}
\includegraphics[width=8cm]{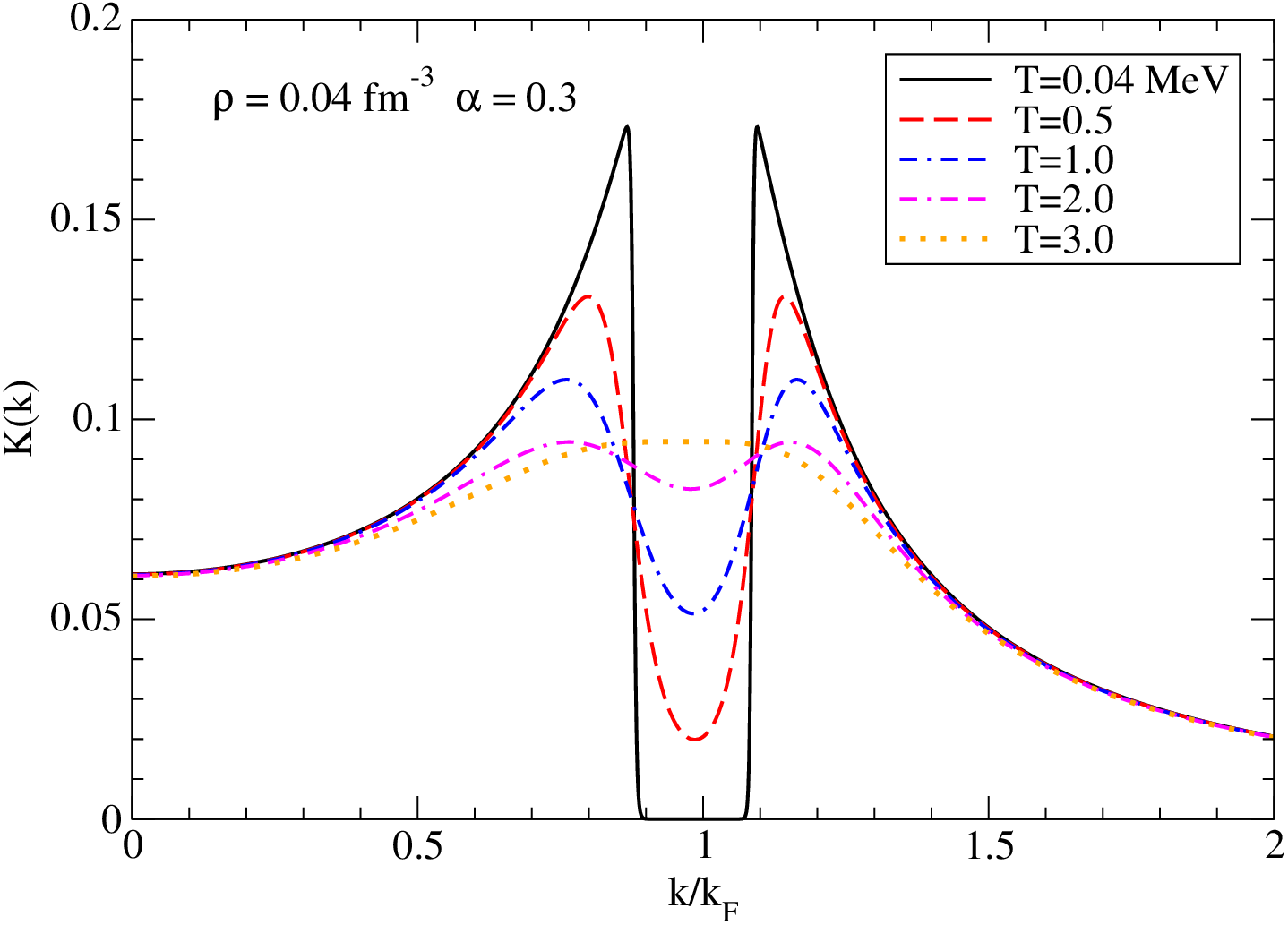}
\caption{(Color online)  Dependence of the kernel $K(k)$ on momentum in
  units of Fermi momentum for fixed $\rho=0.04$ fm$^{-3}$, 
  $\alpha=0.3$, and various temperature indicated in the plot.  }
\label{fig:kernel_temperature}
\end{center}
\end{figure}
\begin{figure}[tb]
\begin{center}
\includegraphics[width=8cm]{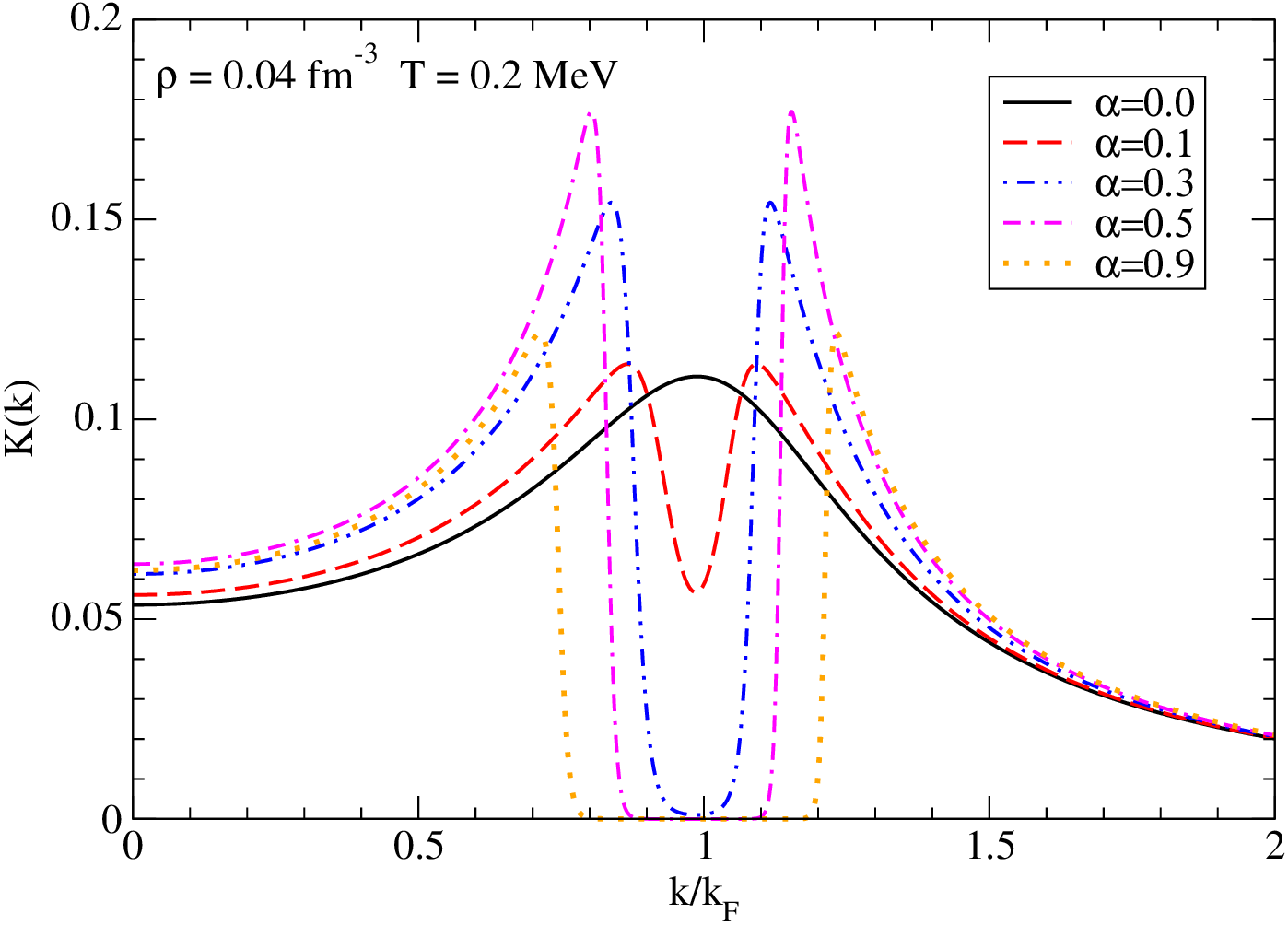}
\caption{(Color online)  Dependence of the kernel $K(k)$ on momentum in
  units of Fermi momentum for fixed $\rho=0.04$ fm$^{-3}$, $T=0.2$ MeV, 
  and various values of asymmetry indicated in the plot. }
\label{fig:kernel_alpha}
\end{center}
\end{figure}
\begin{figure}[tb]
\begin{center}
\includegraphics[width=8cm]{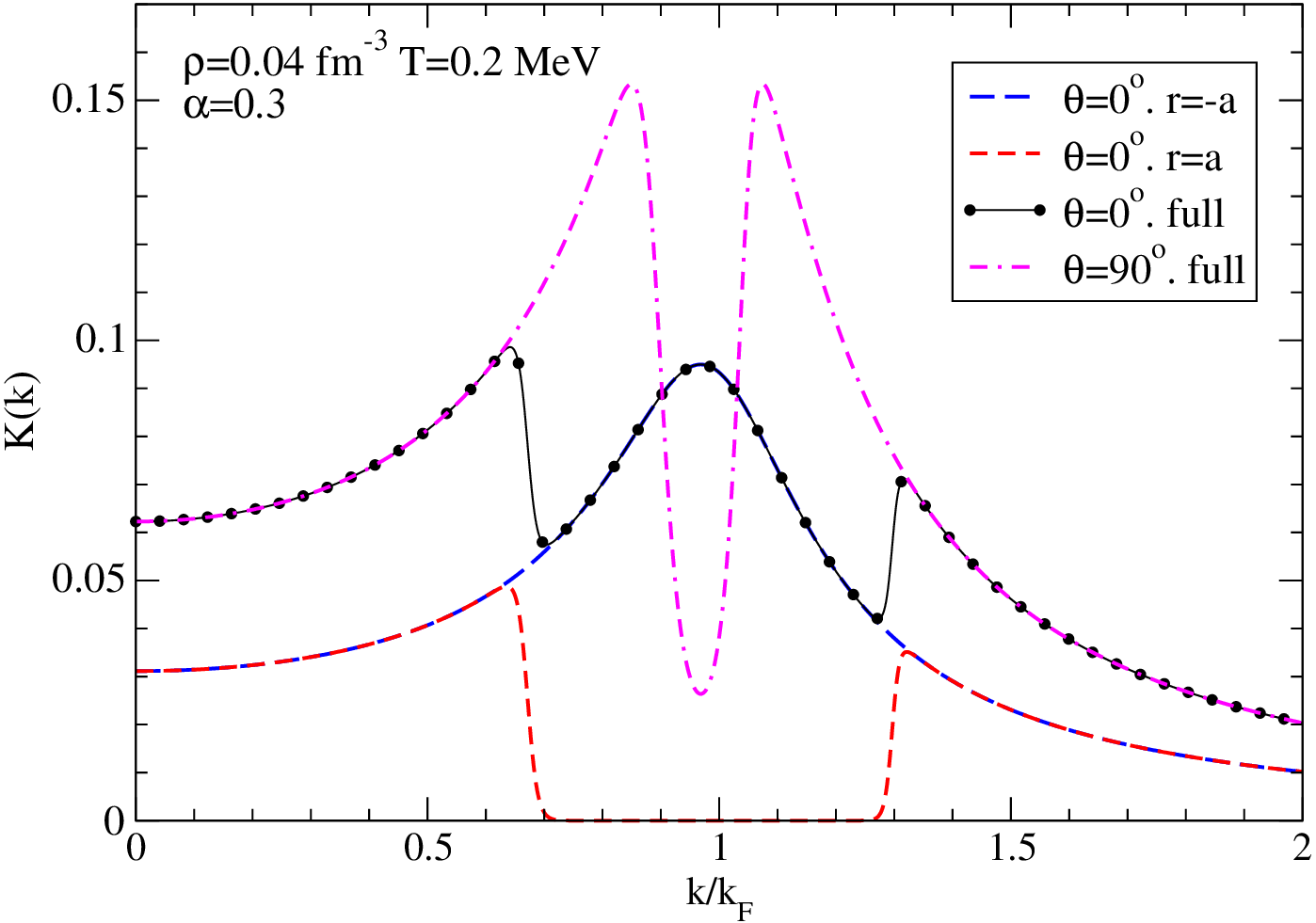}
\caption{(Color online) Dependence of the kernel $K(k)$ on momentum in
  units of Fermi momentum at fixed $\rho=0.04$~fm$^{-3}$, $T=0.2$~MeV,
  and $\alpha=0.3$ for the LOFF phase, where $\theta$ is the angle
  formed by the c. m.  and relative momenta in Eq.~\eqref{eq:kernel}. In
  the case $\theta= 0^o$ the full result (solid-filled-circle line) is
  decomposed into components with $r=-a$ (long dashed line) and $r=a$
  (dash-dotted line).  }
\label{fig:kernel_phases}
\end{center}
\end{figure}

\subsection{The kernel of the gap equations}
\label{subsec:kernel}

The first intrinsic quantity chosen for detailed study is the kernel of the gap equation, 
\bea 
\label{eq:kernel}
K(k,\theta) &\equiv& \sum_{a,r}\frac{P^{a}_r}{4\sqrt{E_{S}(k)^2+\Delta^2(k,Q)}}
.
\eea
This kernel is proportional to the imaginary part of the retarded
anomalous propagator and the Pauli operator represented by $P_r^a =
1-2f(E^a_r)$.  Physically, $K(k)$ can be interpreted as the wave
function of the Cooper pairs, because it obeys a Schr\"odinger-type
eigenvalue equation in the limit of extremely strong coupling. Note
that at Fermi surface $E_S$ vanishes.  The ranges of momenta which
contribute substantially to the gap equation in different regimes of
the phase diagram can be identified from
Figs.~\ref{fig:kernel_density}-\ref{fig:kernel_phases}.  We now
discuss the insights that can be gained from these figures in some
detail.

Figure \ref{fig:kernel_density} shows the function $K(k)$ at constant
temperature and asymmetry for various densities. The high densities
correspond to the BCS regime, and the low densities to the BEC regime,
allowing us to follow the evolution of this function through the
BCS-BEC crossover.  In the BCS regime, $K(k)$ has two sharp maxima
which are separated by a depression of width ∼ $\delta \mu$ around the
Fermi momentum.  Referring to the discussion of occupation numbers in
Sec.~\ref{subsec:numbers} below, this feature originates from the
Pauli operator.  Because of their strong localization in momentum
space, the Cooper pairs have an intrinsic structure that is broad in
real space, implying a large coherence length. This is characteristic
of the BCS regime. The picture is reversed in the strong-coupling
(low-density) limit, where $K(k)$ is a broad function of momentum,
corresponding to the presence of bound states (deuterons), which are
well-localized in real space. This is characteristic of the BEC
regime. In addition, as the density decreases, the lower peak moves
toward $k = 0$, owing to the fact that $\bar\mu$ changes its sign from
positive to negative at the transition from the BCS to the BEC
regime. As a consequence, the prefactor of the Pauli operator $P^r_a $
peaks at $k = 0$ in the BEC regime, rather than at the Fermi surface
as in the BCS regime.

Figure \ref{fig:kernel_temperature} shows the function $K(k)$ for
various temperatures, now at constant asymmetry and constant density,
such that the system is situated in the BCS regime.  At low
temperatures, $K(k)$ is seen to have two maxima separated by a
depression around the Fermi momentum, as already discussed above.
Increasing the temperature smears out the structures characteristic of
the low-temperature case, owing to temperature-induced blurring of the
Fermi surface.  Close to $T_c$, the temperature effects dominate over
the effects of asymmetry. Consequently, the double-peak structure
disappears and the isospin asymmetry does not affect the properties of
the condensate.

Figure~\ref{fig:kernel_alpha} shows the function $K(k)$ for various
asymmetries at constant temperature and the same density as above
(thus again implying the BCS regime).  We can now follow how the
double peak-structure builds up as the asymmetry is increased. Because
the width of the depression is proportional to $\delta\mu$, it
increases with increasing isospin asymmetry, a behavior consistent
with the facts that the Fermi surfaces of neutrons and protons are
pulled apart by the isospin asymmetry and that in the BCS regime the
available phase space is constrained to the vicinity of the
corresponding Fermi surface.

Finally, in Fig.~\ref{fig:kernel_phases} we show $K(k)$ for fixed
values of temperature, asymmetry, and density for the LOFF phase at
two values of the angle formed by the relative and c. m.
momenta, as defined in Eq.~\eqref{eq:kernel}.  It is seen from the
figure that in the orthogonal case ($\theta = 90^{\rm o}$) the
double-peak structure present in the BCS phase remains, although the
effects of asymmetry are weaker compared to the BCS case. This is
easily understood by noting that $E_A = 0$ for $\theta = 90^{\rm
  o}$, therefore finite momentum induces only a shift in the energy
origin according to $\bar\mu \to \bar \mu - Q^2/8m^*$.  
The case $\theta = 0^{\rm o}$ exposes an interesting feature of
the LOFF phase: For a range of orientations of the c. m. momentum of
Cooper pairs ($\theta \sim 0^{\rm o}$), the effects of asymmetry are
mitigated and the kenel obtains a maximum at $k/k_F = 1$, which is a
combination of the contribution from $r=-a$, which acts to enhance the
pairing correlations in the vicinity of the Fermi surface, and the $r =
a$ contribution which vanishes in this region.

\subsection{The Cooper-pair wave function across the BCS-BEC phase
  transition}
\label{subsec:CooperWave}

The transition to the BEC regime of strongly coupled neutron-proton
pairs, which are asymptotically identical with deuterons, occurs at
low densities.  The criterion for the transition from BCS to BEC is
that either the average chemical potential $\bar \mu$ changes its sign
from positive to negative values, or the coherence length $\xi$ of a
Cooper pair becomes comparable to the interparticle distance, i.e.,
$\xi$ becomes of order $d\sim \rho^{-1/3}$. (In the BCS regime 
$\xi \gg d$, whereas in the BEC regime $\xi \ll d$.)

The coherence length can be related to the root mean square of the
Cooper-pair wave function, as we show below.  The wave function of a
Cooper pair is defined in terms of the kernel of the gap equation
according to 
\bea \label{eq:Psi} \Psi(\vecr) = \sqrt{N} \int
\frac{d^3p}{(2\pi)^3}
[K(\vecp,\Delta)-K(\vecp,0)]e^{i\vecp\cdot\vecr}, 
\eea 
where $N$ is a constant determined by the normalization condition 
\be
N\int d^3r \vert \Psi(\vecr)\vert^2 = 1.  
\ee
In Eq.~(\ref{eq:Psi}) we subtract from the kernel its value 
$K(\vecp,0)$ in the normal state 
to regularize the integral, which is
otherwise divergent. Cut-off regularization of this strongly
oscillating integral is not appropriate.  The mean-square radius of a
Cooper pair is defined via the 
second moment of the probability density,
\be \langle r^2\rangle = \int d^3r\, r^2 \vert
\Psi(\vecr)\vert^2.  \ee 
The coherence length, i.e., the spatial extension of a Cooper pair, is
then defined as
\be
\xi_{\rm rms} = \sqrt{\langle
  r^2\rangle}.
\ee 
Thus the change in the coherence length is related to the change of
the condensate wave function across the BCS-BEC crossover.  The
regimes of strong and weak coupling can be identified by comparing the
coherence length to the mean interparticle distance $d = (3/4\pi
\rho)^{1/3}$.  In the BCS regime the coherence length is given by the
well-known analytical formula
\bea \xi_a = \frac{\hbar^2
  k_F}{\pi m^* \Delta}.  \eea
Table I lists the analytical and root-mean-square values of the
coherence length for several densities and temperatures, chosen to
represent the different regimes WCR, ICR, and SCR, together the
corresponding values of the mean interparticle distance.  It is seen
that in the case of neutron-proton pairing, one of the criteria for
the BCS-BEC transition is fulfilled, namely, the mean distance between
the pairs becomes larger than the coherence length of the superfluid
as one goes from WCR to SCR. We have verified that the average
chemical potential changes its sign accordingly, so that the second
criterion is fulfilled as well.
\begin{table}
\begin{tabular}{ccccccc}
\hline
 & log$_{10}\left(\frac{\rho}{\rho_0}\right)$ & $k_F$[fm$^{-1}$] &$T$ [MeV] & $d$
[fm] & $\xi_{\rm rms}$ [fm] & $\xi_{a}$ [fm] \\
\hline\hline
WCR & $-0.5$ &0.91& 0.5 & 1.68 & 3.17 & 1.41 \\
ICR & $-1.5$  & 0.42& 0.5 & 3.61 & 0.94 & 1.25 \\
SCR & $-2.5$ & 0.20 & 0.2 & 7.79 & 0.57 & 1.79 \\
\hline\\
\end{tabular}
\caption{
  For each of the three regimes of coupling strength, 
  corresponding values are presented for the density 
  $\rho$ (in units of nuclear saturation density 
  $\rho_0 = 0.16$ fm$^{-3}$), Fermi momentum $k_F$, temperature $T$, 
  interparticle distance $d$, 
  and coherence parameters $\xi_{\rm rms}$ and $\xi_{a}$. 
The values of the gap and effective mass (in units of bare mass)  at
 $\alpha = 0$ in these three regimes are 9.39, 4.50, 
 1.44 MeV and 0.903, 0.989, 0.999, respectively. 
  In the regime WCR, the LOFF phase is found in the vicinity of 
  asymmetry  $\alpha = 0.49$, for which  $\Delta = 1.27$ MeV and 
  $Q = 0.4$ fm$^{-1}$. 
}
\end{table}
\begin{figure}[tb]
\begin{center}
\includegraphics[width=8cm]{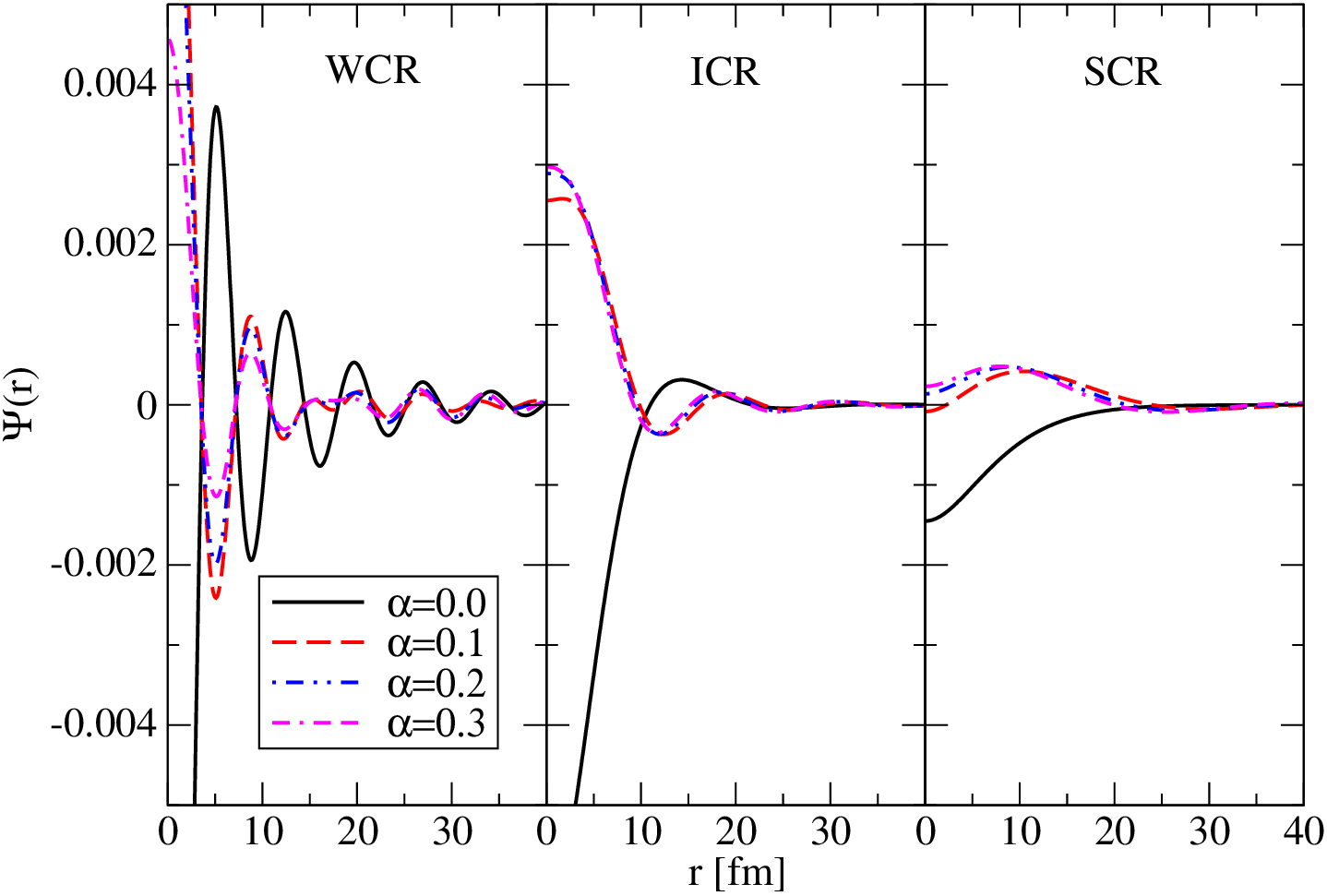}
\caption{(Color online) Dependence of $\Psi(r)$ on $r$ for the three
  coupling regimes and various values of asymmetry (see Table I for
  values of density and temperature).  }
\label{fig:Psi}
\end{center}
\end{figure}
\begin{figure}[tb]
\begin{center}
\includegraphics[width=8cm]{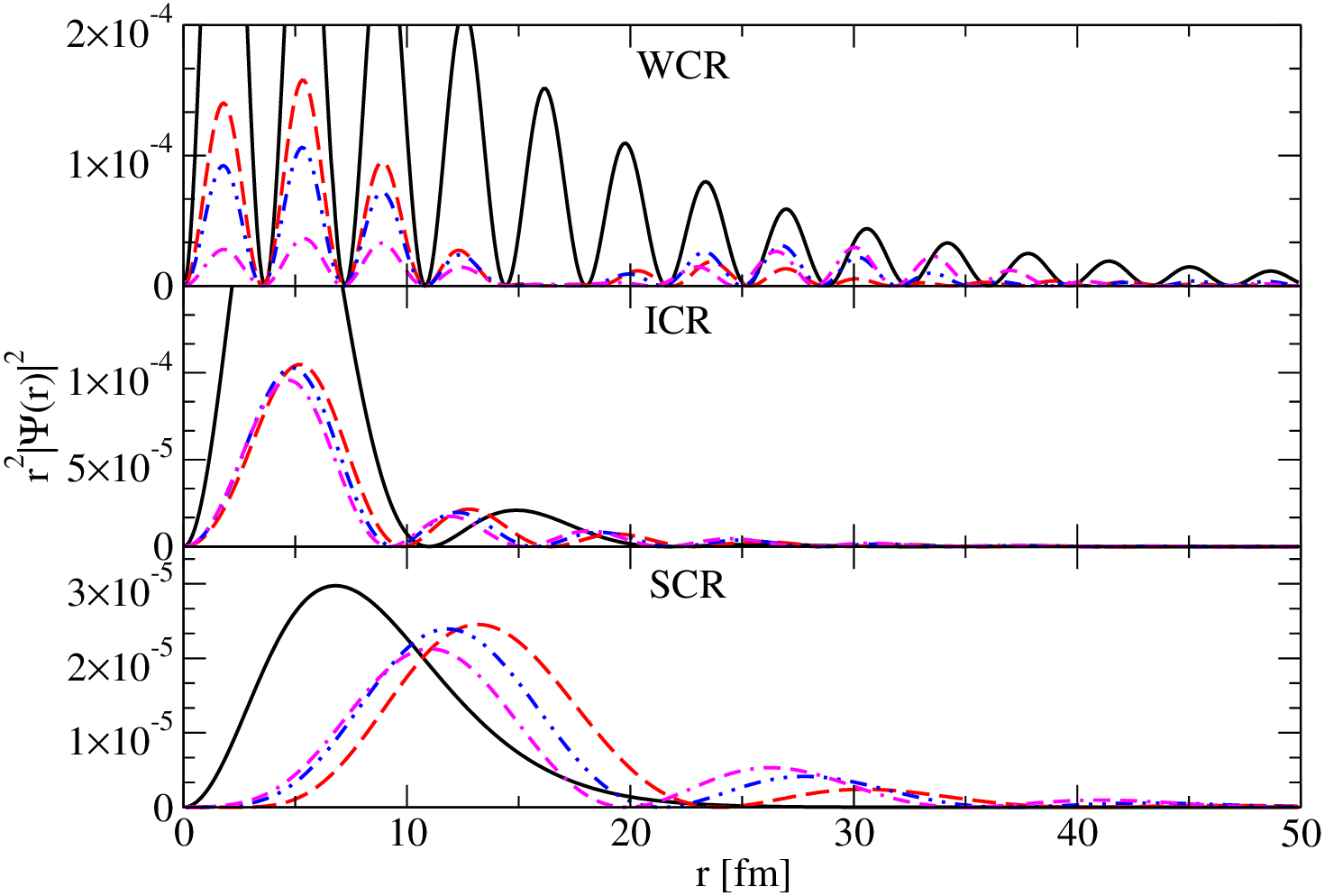}
\caption{(Color online) Dependence of $r^2|\Psi(r)|^2$ on $r$ for the
  three coupling regimes.  Conventions are the same as in
  Fig.~\ref{fig:Psi}.}
\label{fig:r_Psi}
\end{center}
\end{figure} 
Figure \ref{fig:Psi} shows the wave function of Cooper pairs as a
function of radial distance across the BCS-BEC crossover for various
densities.  In weak coupling, the wave function has a well-defined
oscillatory form that extends over many periods of the interparticle
distance. Such a state conforms to the familiar BCS picture, in which
the spatial correlations are characterized by scales that are much
larger than the interparticle distance. For intermediate and strong
coupling the wave function is increasingly concentrated at the origin
with at most a few periods of oscillation.  The strong-coupling limit
corresponds to pairs that are well localized in space within a small
radius. This regime clearly has BEC character, with the pair
correlations extending only over distances comparable to the
interparticle distance. It is seen that in weak coupling the wave
function is almost independent of the asymmetry, whereas in strong
coupling this dependence is substantial.
\begin{figure}[tb]
\begin{center}
\includegraphics[width=8cm]{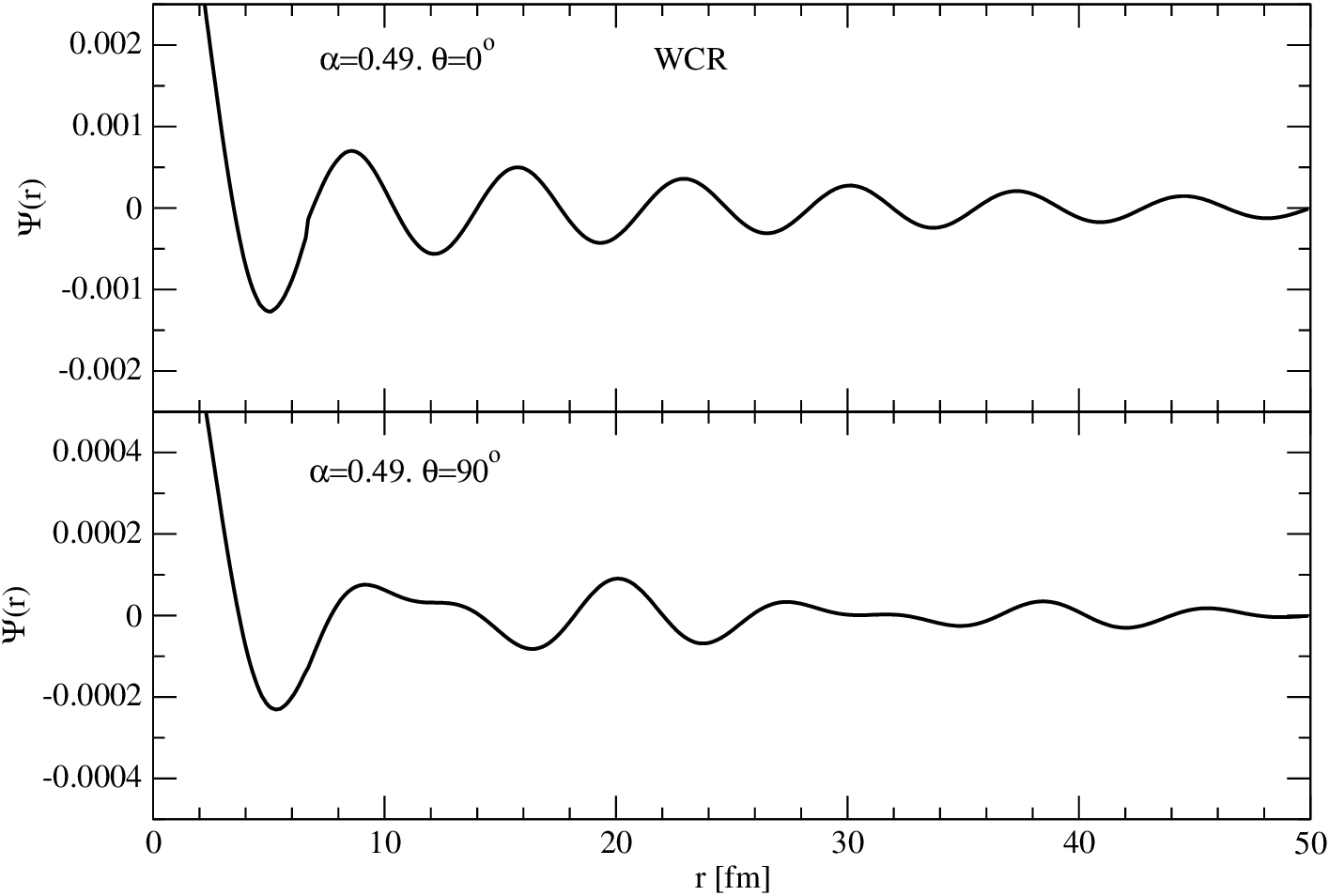}
\caption{Dependence of $\Psi(r)$ on $r$ in the regime WCR for two
  different angles $\theta$ for asymmetry $\alpha = 0.49$ at which the
  LOFF phase is the ground state.  }
\label{fig:Psi_loff1}
\end{center}
\end{figure}
\begin{figure}[tb]
\begin{center}
\includegraphics[width=8cm]{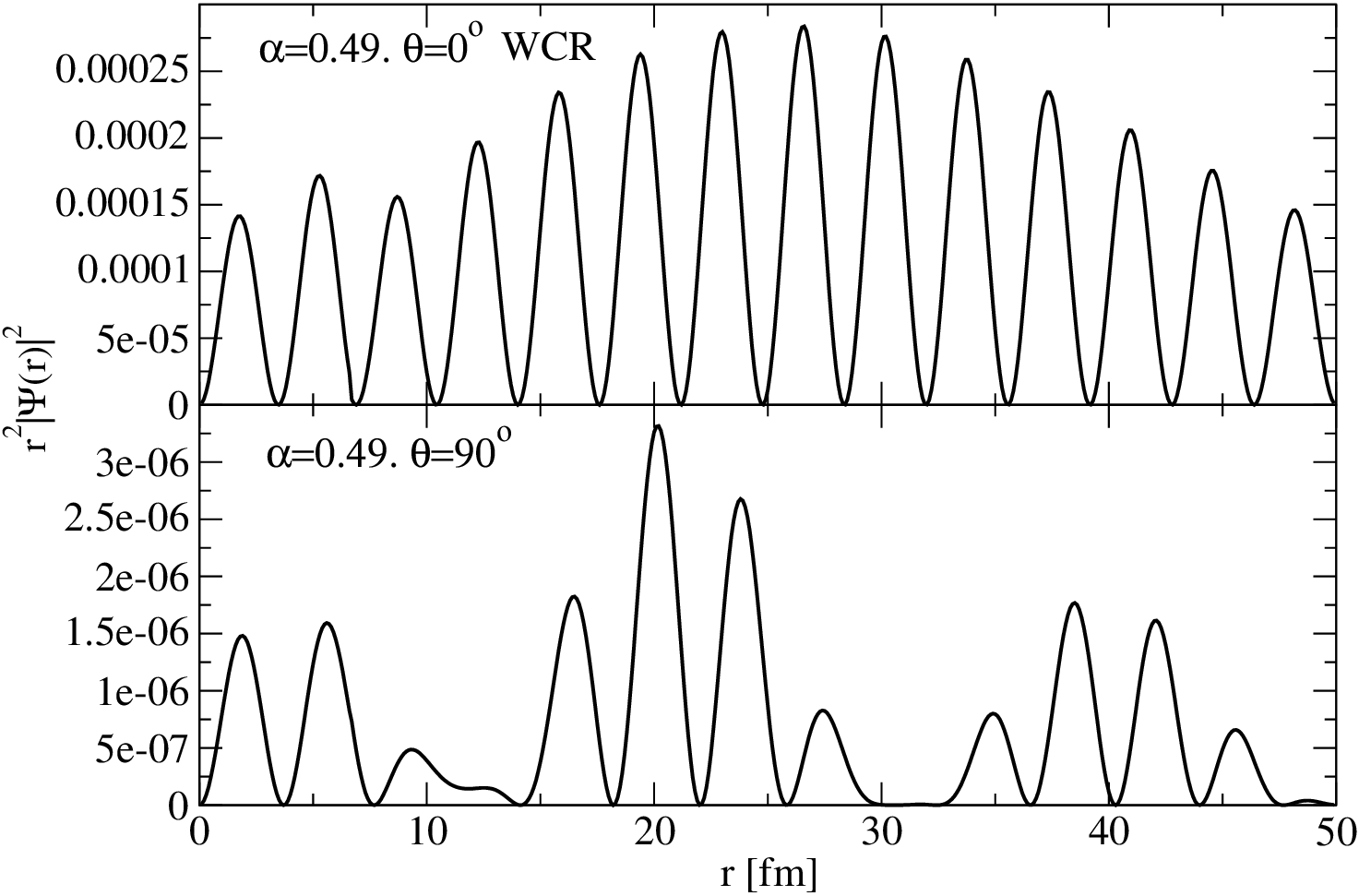}
\caption{Dependence of $r^2|\Psi(r)|^2$ on $r$ in the regime WCR for
  two different angles $\theta$ for asymmetry $\alpha = 0.49$ at which
  the LOFF phase is the ground state.  }
\label{fig:Psi_loff2}
\end{center}
\end{figure}
Figure~\ref{fig:r_Psi}, complementary to Fig.~\ref{fig:Psi}, displays
the quantity $r^2\vert \Psi(\vecr)\vert^2$.  The spatial correlation
in the regime SCR is dominated by a single peak corresponding to a
tightly bound state close to the origin.  The existence of residual
oscillations indicates that there is no unique bound state formed at
such coupling, but the tendency towards its formation is clearly
seen. An oscillatory structure appears in the ICR as a fingerprint of
the transition from BEC the to the BCS regime. In the WCR we observe
oscillations over many periods, i.e., over large distances, indicative
of the coherent BCS state.  At low and high asymmetries the
strong-coupling peaks are well defined, whereas at intermediate
asymmetries the weight of the function is distributed among several
peaks.

Figure~\ref{fig:Psi_loff1} and \ref{fig:Psi_loff2} demonstrates the
same quantities $\Psi(\vecr)$ and $r^2\vert \Psi(\vecr)\vert^2$ for
the case of the LOFF phase computed at the WCR point of the phase
diagram (as specified in Table I).  At this point the LOFF phase is
the ground state of the matter at asymmetry $\alpha = 0.49$
($\delta\mu=6.45$ MeV), where $\Delta = 1.27$ MeV and $Q = 0.4$
fm$^{-1}$. For slightly lower asymmetries ($\alpha \le 0.48$) the
system is in the PS phase, whereas for $\alpha > 0.5$ the gap is
vanishingly small, the system being in the normal state.
 In the case $\theta=0^{\rm o}$ the perfect oscillatory
  behavior seen in $\Psi(\vecr)$ in the BCS case is replicated, as in
  this case the finite momentum of the condensate does not contribute
  to the spectrum of the Cooper pairs. In the case $\theta=90^o$
  $\Psi(\vecr)$ is distorted in the LOFF phase by the presence of a
  second oscillatory mode with the period $2\pi/Q$ in addition to the
  first mode, with the period $2\pi/k_F$. The additional periodic structure
  is more pronounced in the quantity $r^2\vert \Psi(\vecr)\vert^2$,
  where the rapid oscillations are modulated with a period $\sim 16$~fm. 
\subsection{Occupation numbers}
\label{subsec:numbers}
\begin{figure}[tb]
\begin{center}
\includegraphics[width=8cm]{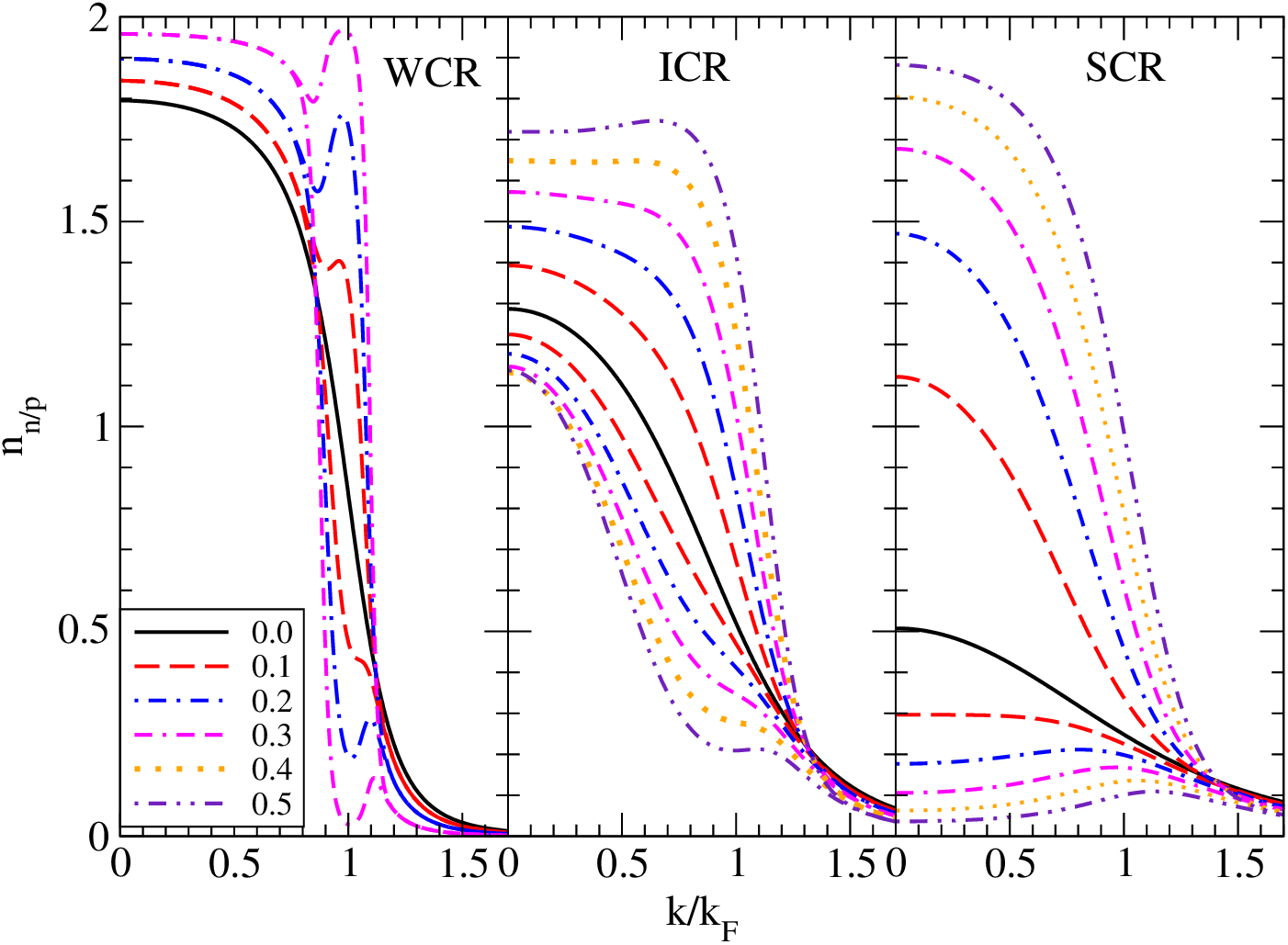}
\caption{(Color online) Dependence of the neutron and proton
  occupation numbers on momentum $k$ (in units of Fermi momentum) for
  the three coupling regimes and various asymmetries indicated in the
  legend.  }
\label{n_p_dens}
\end{center}
\end{figure}

\begin{figure}[tb]
\begin{center}
\includegraphics[width=8cm]{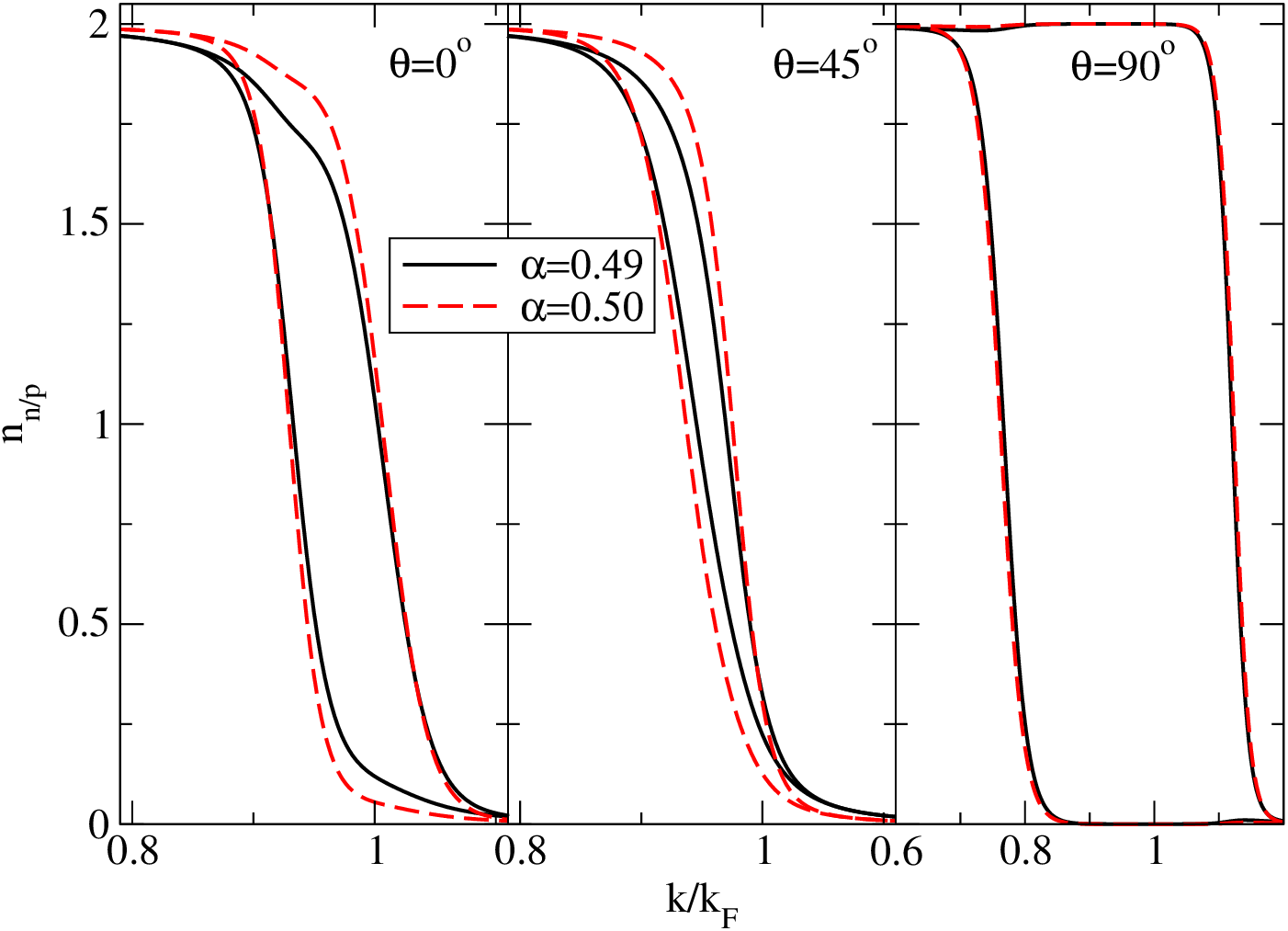}
\caption{ (Color online) Dependence of the neutron and proton
  occupation numbers on momentum $k$ (in units of Fermi momentum) in
  the WCR for two asymmetries where the LOFF phase is the ground
  state. The three angles indicated refer to the neutron occupation
  numbers. The proton occupation numbers are plotted for angles
  $180^{\rm o}-\theta$. }
\label{fig:n_loff_weak}
\end{center}
\end{figure}
The integrand of Eq.~\eqref{eq:densities} defines the occupation
numbers $n_{n/p}(k)$ of the neutrons and protons. These quantities are
shown in different coupling regimes of the BCS-BEC crossover in
Fig.~\ref{n_p_dens}.  In the WCR (leftmost panel) the occupation
numbers of protons exhibit a ``breach'' \cite{2003PhRvL..91c2001G}
or ``blocking region'' for
large asymmetries, i.e., the minority component is entirely expelled
from the blocking region ($n_p=0$), while the majority component is
maximally occupied ($n_n/2=1$).  In the small-$\alpha$ limit the
occupation numbers are clearly fermionic (with some diffuseness owing to
the temperature) in that all single-particle states below a certain
mode (the Fermi momentum at $T=0$) are almost filled, while all states
above are nearly empty.  We have verified that in the high-temperature
limit the breach is filled in, the occupation numbers becoming smooth
functions of momentum; consequently the low-momentum modes are less
populated.

In the ICR (middle panel) the fermionic nature of the occupation
numbers is lost. The low-momentum modes are not fully populated and,
accordingly, high-momentum modes are more heavily occupied.  A Fermi
surface cannot be identified because of the smooth population of the
modes.  Moreover, a breach no longer appears for the parameters
chosen.  It is also to be noted that for large asymmetries $\alpha \ge
0.4$, the momentum dependence of the occupation numbers becomes
non-monotonic; for the minority component this is a precursor of the
change in the topology of the Fermi surface
under increase of coupling strength.

The SCR (rightmost panel) can be identified with the BEC phase of
strongly coupled pairs. At large asymmetries the distribution of the
minority component undergoes a topological change.  First there
develops an empty strip within the distribution function, which is
reorganized at larger asymmetries into a distribution in which the
modes are populated starting from a certain nonzero value.  Thus, the
Fermi sphere occupied by the minority component in the weakly coupled
BCS limit evolves into a shallow shell structure in the strongly
coupled Bose-Einstein-condensed limit. This behavior was already
revealed in the case of the $\SD$ condensate in
Ref.~\cite{2001PhRvC..64f4314L}.

Figure~\ref{fig:n_loff_weak} depicts the occupation numbers in the WCR
at asymmetries corresponding to a LOFF-phase ground state for three
fixed angles $\theta = 0^{\rm o}$, $45^{\rm o}$, and $90^{\rm o}$.  In
the case $\theta = 90^{\rm o}$ we have $E_A = 0$, and the LOFF
spectrum differs from the asymmetrical BCS spectrum only by a shift in
the energy origin, $\bar\mu \to \bar \mu - Q^2/8m^*$.  Therefore the
occupation numbers do not depart qualitatively
from their BCS behavior; moreover, the
``breach'' is clearly seen.  For 
$\theta = 45^{\rm o}$ the difference between the occupation numbers
disappears, i.e., the superconductor behaves as if it were isospin
symmetric. This result follows from the fact that the nonzero c. m.
momentum of the LOFF phase compensates for the mismatch of the Fermi
spheres and restores the coherence needed for pairing. In the case
$\theta = 0^{\rm o}$ the effect of $E_A$ attains its maximal value,
but the occupation numbers are intermediate between those of the two
cases previously addressed. This is attributable to the fact that the overlap
between the spectra of neutron and proton quasiparticle branches is
beter for $\theta = 45^{\rm o}$ than for $\theta = 0^{\rm
  o}$, in which case the quasiparticle spectra ``overshoot'' the
optimal overlap (see the discussion in the following section).

\subsection{Quasiparticle spectra}
\label{subsec:DR}

\begin{figure}[tb]
\begin{center}
\includegraphics[width=8cm]{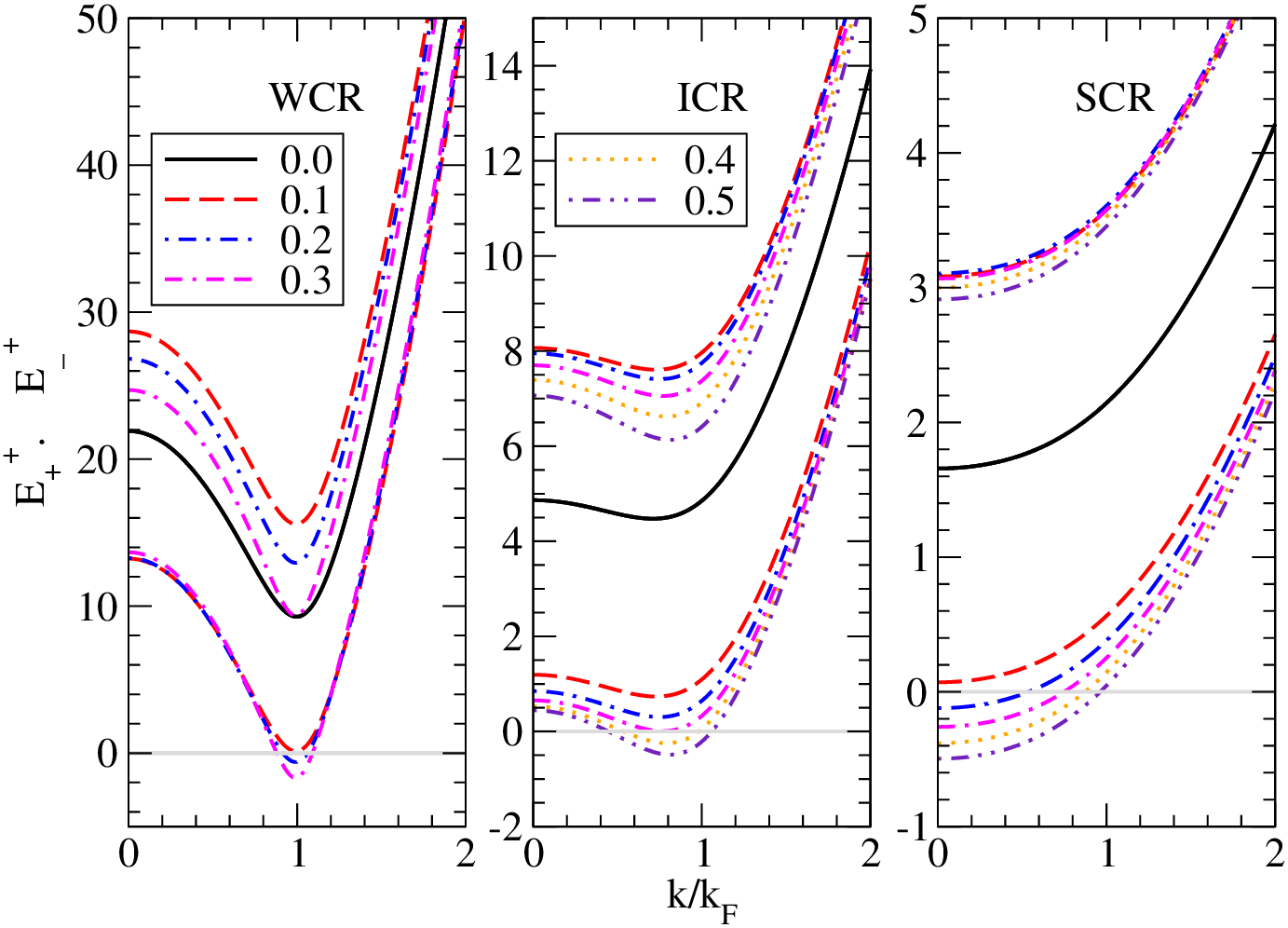}
\caption{(Color online) Dispersion relations for quasiparticle spectra
  in the case of the BCS condensate, as functions of momentum in units
  of Fermi momentum.  For each asymmetry, the upper branch corresponds
  to $E^+_+$, and the lower to the $E^+_-$ solution.   }
\label{fig:spectra_bcs}
\end{center}
\end{figure}

Finally, let us consider the dispersion relations for quasiparticle
excitations about the $\SD$ condensate. We first examine in some
detail the spectra $E^a_{\pm}$ in the BCS case defined in
Eq.~\eqref{eq:QP_spectra}, which are then independent of the sign of
$a$ and we take $a = +$.  These are shown in
Fig.~\ref{fig:spectra_bcs} for the three coupling regimes of
interest. In the isospin-symmetric BCS case, the dispersion relation
has a minimum at $E^+_{+} = E^+_{-} = \Delta$ for $k= k_F$. For finite
asymmetries one has $E^+_{\pm} = \sqrt{E_S^2+\Delta^2} \pm \delta\mu$;
hence the minima of the dispersion relations of neutron and proton
quasiparticles are given by an asymmetry-dependent gap value modified
by the shift in chemical potential, i.e.,
$\Delta(\alpha)\pm\delta\mu$. For protons this leads to a gapless
spectrum, which does not require a finite minimum energy for
excitation of two modes (say $k_1$ and $k_2$) for which the dispersion
relation intersects the zero-energy axis. This phenomenon is well
known as {\it gapless superconductivity}. The momentum interval
$k_1\le k\le k_2$ corresponds to the interval in Fig.~\ref{n_p_dens}
where the occupation numbers of majority and minority components
separate and the ``breach'' in the occupation of the minority
component becomes prominent.

Consider now the SCR, in which case we are dealing with a gas of
deuterons and free neutrons. In the symmetrical limit (i.e.\ when only
deuterons are present), the dispersion relation has a minimum at the
origin that corresponds to the (average) chemical potential, which
asymptotically approaches half the binding energy of a deuteron in
vacuum~\cite{2001PhRvC..64f4314L}. The effect of asymmetry is to shift
the average chemical potential downwards and to introduce the
separation $\delta\mu$ in the quasiparticle spectra.

Because the minimum is now at the origin, there is only one mode for
which the dispersion relation crosses zero at a finite $k$. The
dispersion relations in the ICR experience a transition from the WCR
to the SCR, such that their key features resemble those of the WCR,
but with a shallower minimum and a larger momentum interval
$[k_1,k_2]$ over which the excitation spectrum becomes gapless.
\begin{figure}[t]
\begin{center}
\includegraphics[width=8cm]{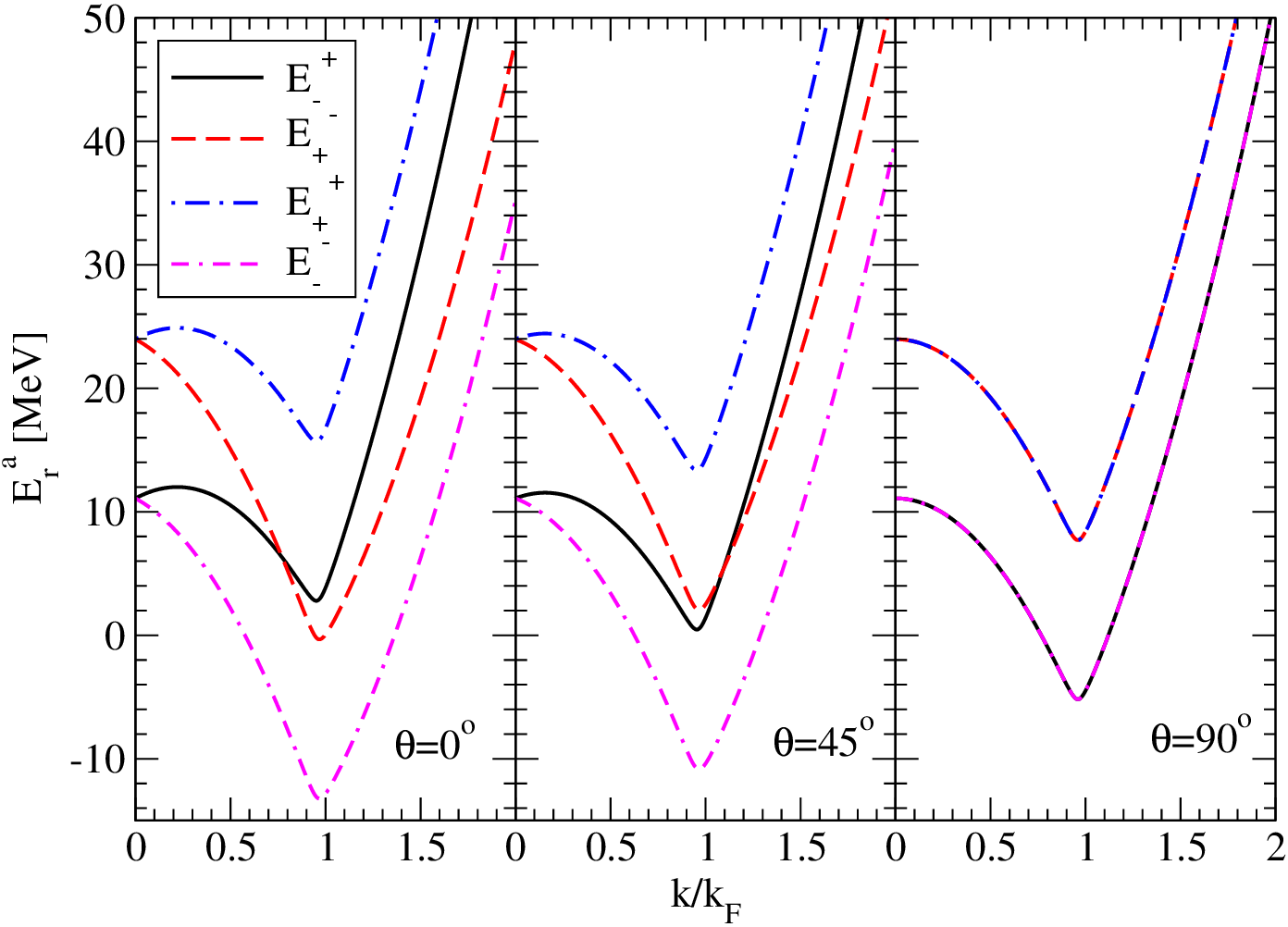}
\caption{(Color online) Dispersion relations for quasiparticle spectra
in the LOFF phase for three angles and $\alpha = 0.49$.  }
\label{fig:spectra_loff}
\end{center}
\end{figure}

The dispersion relations for quasiparticles in the LOFF phase for
special angles $\theta$ are shown in Fig.~\ref{fig:spectra_loff} in
the WCR and for a $\alpha$ values corresponding to the LOFF phase as
ground state. In this case, we show all four branches of quasiparticle
spectrum. Consistent with the earlier discussion of
Fig.~\ref{fig:n_loff_weak} for $\theta =90^{\rm o}$, the LOFF phase
resembles the BCS phase and there is a large mismatch between the
spectra of protons and neutrons. In this case the branches $a = +$ and
$a=-$ are degenerate. For other angles we see again that the nonzero
c. m. momentum mitigates the asymmetry and brings the quasiparticle
spectra closer together, i.e., the LOFF phase resembles the
symmetrical BCS phase for the two branches with $a \neq r$ for $\theta
< 90^o$.  This is
particularly clear for $\theta = 45^{\rm o}$, in which case two of the
four dispersion relations coincide in the vicinity of the Fermi
momentum. It is clear that the optimal mitigation of the isospin
mismatch by the finite moment does not need to be for $\theta = 0^o$,
but can occur at some angle $0^o\le \theta\le 90^o$; it is seen that for
$\theta = 0^o$ the branches cross and, hence, ``overshoot'' the optimal
compensation.

\begin{figure}[tb]
\begin{center}
\includegraphics[width=7cm]{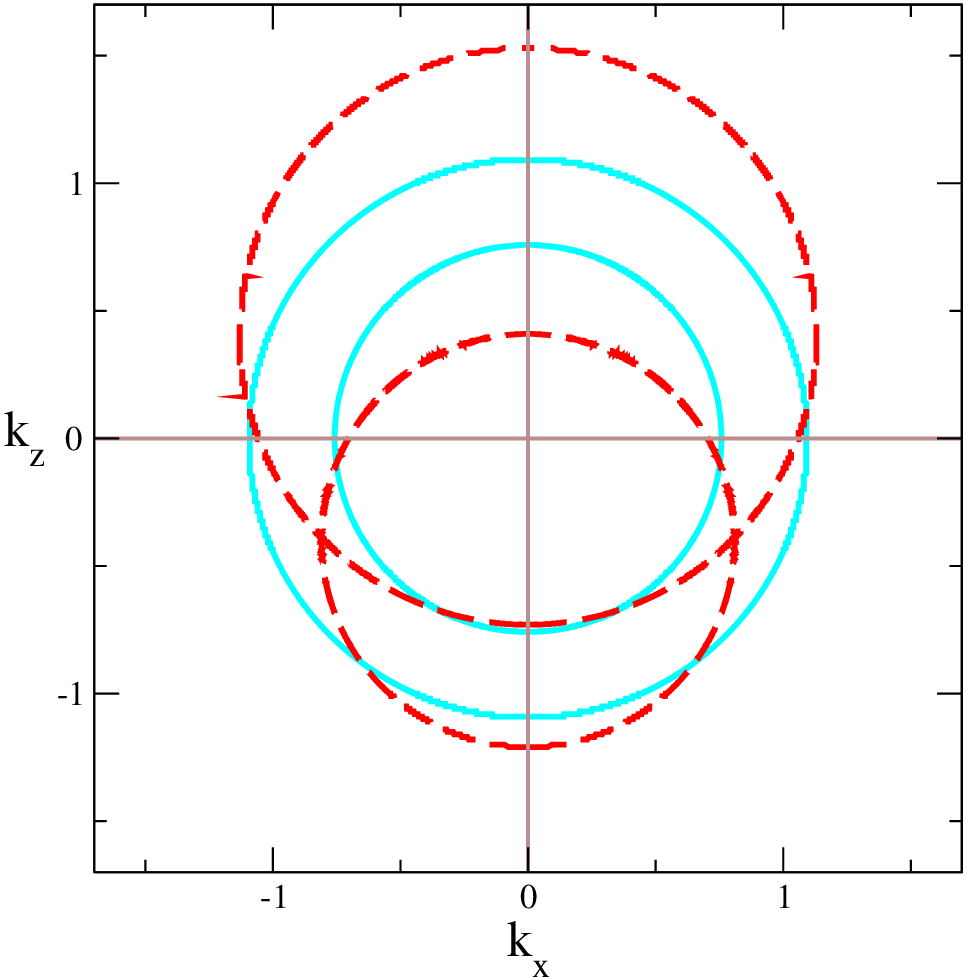}
\caption{(Color online) Illustration of Fermi surfaces in the
  asymmetrical BCS state (solid lines) and LOFF phase (dashed lines).
  The LOFF phase is characterized by the following values of
  parameters: $\alpha = 0.49$, $\delta\mu=6.45$~MeV, $\Delta =
  1.27$~MeV and $Q = 0.4$~fm$^{-1}$.  }
\label{fig:spheres}
\end{center}
\end{figure}

The restoration of the coherence (Fermi-surface overlap) in the LOFF
phase can be illustrated by looking at the solutions of
$\epsilon^{\pm}_{n/p,\uparrow/\downarrow} = 0$ [see
Eq.~\eqref{eq:norm_spec}] which define the Fermi-surface in the limit
$\Delta\to 0$ but $Q\neq 0$. These are illustrated in
Fig.~\ref{fig:spheres} in two cases, $Q = 0$ and $Q\neq 0$. In the
first case the Fermi surfaces are concentric spheres which have no
intersection. In the second case the non-zero c. m. leads to an
intersection of the Fermi-spheres; in these regions of intersection
the pair correlations are restored to the magnitude characteristic to
the BCS phase. Of course, the c. m. momentum costs positive kinetic
energy, which must be smaller than the negative condensation energy
for LOFF phase to be stable. 

\section{Conclusion}
\label{sec:conclude}

Low-density nuclear matter is predicted to feature a rich phase
diagram at low temperatures and nonzero isospin asymmetry. The phase
diagram contains at least the following phases: the translationally
and rotationally symmetric, but isospin-asymmetrical BCS phase, the
BEC phase containing neutron-proton dimers, the current-carrying LOFF
phase, and associated phase-separated phases.

Our analysis of these phases can be summarized as follows.
\begin{itemize}
\item The phase diagram of nuclear matter composed of these phases has
two tri-critical points in general, one of which is a Lifshitz
point. These can combine in a tetra-critical point for a special
combination of density, temperature, and isospin asymmetry. The phase
diagram contains two types of crossovers from the asymmetrical BCS
phase to the BEC of deuterons and an embedded neutron gas: a
transition between the homogeneous BCS-BEC phases at relatively high
temperatures and between the heterogeneous BCS-BEC phases at low
temperatures. We have shown that the LOFF phase exists only in a
narrow strip in the high-density, low-temperature domain and at
nonzero asymmetries.

\item The crossovers of BCS-BEC type are smooth and are characterized
  by lines in the temperature-density plane that are insensitive to
  the isospin asymmetry. These lines were obtained by examining the
  sign of the average chemical potential.

\item Detailed analysis of key intrinsic quantities, including the
  kernel of the gap equation along with the Cooper-pair wave function
  and its probability density, clearly establishes that in the BCS
  limit one deals with a coherent state, whose wave function
  oscillates over many periods with a wavelength characterized by the
  inverse Fermi momentum $k_F^{-1}$. In the opposite limit the wave
  function is well-localized around the origin, indicating that one is
  then dealing with a Bose condensate of strongly bound states, namely
  deuterons.

\item The analysis of the kernel of the wave function, the occupation
  probabilities of neutrons and protons, and the quasiparticle
  dispersion relations demonstrates the prominent role played by the
  Pauli-blocking region (called ``the breach'')
  \cite{2003PhRvL..91c2001G} that appears in these quantities.  In the
  BCS phase and the low-temperature limit of the WCR, the blocking
  region embraces modes in the range $k_1\le k\le k_2$ around the
  Fermi surface.  In this modal region, it has been found that (a) the
  minor constituents (protons) are extinct; (b) there are no
  contributions to the kernel of the gap equation from these modes;
  and (c) the end of points of this region correspond to the onset of
  gapless modes that can be excited without any energy cost.  The LOFF
  phase appearing in this regime substantially mitigates the blocking
  mechanism by allowing for nonzero c. m.  momentum of the
  condensate. As a consequence, all the intrinsic quantities studied
  are much closer to those of the isospin-symmetric BCS state.

\item We have traced the evolution of the targeted intrinsic
  properties into the SCR as the system
  crosses over from the BCS condensate to a BEC of deuterons plus a
  neutron gas. In the SCR the long-range coherence of the condensate
  is lost. The dispersion relations change their form from a spectrum
  having a minimum at the Fermi surface to a spectrum that is minimal
  at $k=0$, as would be expected for a BEC, independent of isospin
  asymmetry.  With increasing isospin asymmetry, the proton dispersion
  relation acquires points with zero excitation energy in this regime.
  The occupation numbers reach a maximum for finite $k$ and reflect a
  change of topology at large asymmetries: The filled ``Fermi sphere''
  becomes an empty ``core.''
\end{itemize}

The present investigation of BCS-BEC crossovers with inclusion of
unconventional phases, such as the LOFF phase and the heterogeneous
phase-separated phase, could be useful in the studies of
spin/flavor-imbalanced fermionic systems in ultracold atomic gases,
for recent studies see, e.g.,
Refs.~\cite{2007AnPhy.322.1790S,2008RvMP...80.1215G,2011JPhCS.321a2028S},
dense quark matter (e.g.,
Refs.~\cite{2009PhRvD..80g4022S,2010PhRvD..82e6006M,2014PhRvD..89c6009M,2010PhRvD..82g6002F,2012NuPhA.875...94K}),
and other related quantum systems.

\acknowledgments

M. S. acknowledges the support by the HGS-HIRe graduate program at
Frankfurt University. A. S. is supported by the Deutsche
Forschungsgemeinschaft (Grant No. SE 1836/3-1) and thanks the
Institute for Nuclear Theory at the University of Washington, Seattle,
for its hospitality, and the Department of Energy for partial support
during the program ``Binary Neutron Star Coalescence as a Fundamental
Physics Laboratory.''  X. G. H.  is supported by Fudan University
Grant No.
EZH1512519 and Shanghai Natural Science Foundation Grant
No. 14ZR1403000.  J. W. C. acknowledges research support from the McDonnell
Center for the Space Sciences and expresses his thanks to Professor
Jos\'e Lu\'is da Silva and his colleagues at Centro de Ci\^encias
Matem\'aticas for gracious hospitality at the University of Madeira.

\end{document}